\documentclass[prr,a4,superscriptaddress,preprintnumbers,twoside,twocolumn,floatfix, longbibliography]{revtex4-2}

\usepackage[table]{xcolor}
\usepackage{import}
\usepackage{stmaryrd}
\usepackage{physics}
\usepackage{color}
\usepackage{bbm}
\usepackage{enumerate}
\usepackage{float}

\usepackage{amsfonts,amssymb} 
\usepackage[export]{adjustbox}
\usepackage{soul,xcolor}

\usepackage{soul}
\usepackage[english]{babel}
\usepackage[utf8]{inputenc}
\usepackage[T1]{fontenc}
\usepackage[colorlinks=true, allcolors=blue]{hyperref}
\hypersetup{
	colorlinks=true,  
	linkcolor=blue,   
	citecolor=blue,   
	urlcolor=blue     
}

\begin{document}

\setstcolor{blue}

\title[ATI]{Enhancing quantum state tomography via resource-efficient attention-based neural networks}

\author{Adriano Macarone Palmieri\footnote{These authors contributed equally.}}
\email{adriano.macarone@icfo.eu}
\altaffiliation{These authors contributed equally.}
\affiliation{ICFO - Institut de Ciències Fotòniques, The Barcelona Institute of Science and Technology, 08860 Castelldefels (Barcelona), Spain}

\author{Guillem Müller-Rigat\footnote{These authors contributed equally.}}
\email{guillem.muller@icfo.eu}
\altaffiliation{These authors contributed equally.}
\affiliation{ICFO - Institut de Ciències Fotòniques, The Barcelona Institute of Science and Technology, 08860 Castelldefels (Barcelona), Spain}

\author{Anubhav Kumar Srivastava}
\affiliation{ICFO - Institut de Ciències Fotòniques, The Barcelona Institute of Science and Technology, 08860 Castelldefels (Barcelona), Spain}

\author{Maciej Lewenstein}
\affiliation{ICFO - Institut de Ciències Fotòniques, The Barcelona Institute of Science and Technology, 08860 Castelldefels (Barcelona), Spain}
\affiliation{ICREA, Pg. Lluis Companys 23, 08010 Barcelona, Spain}

\author{Grzegorz Rajchel-Mieldzioć}
\affiliation{ICFO - Institut de Ciències Fotòniques, The Barcelona Institute of Science and Technology, 08860 Castelldefels (Barcelona), Spain}
\affiliation{NASK National Research Institute, ul. Kolska 12 01-045 Warszawa, Poland}

\author{Marcin Płodzień}\email{marcin.plodzien@icfo.eu}
\affiliation{ICFO - Institut de Ciències Fotòniques, The Barcelona Institute of Science and Technology, 08860 Castelldefels (Barcelona), Spain}

\begin{abstract}

In this work, we propose a method for denoising experimental density matrices that combines standard quantum state tomography with an attention-based neural network architecture. The algorithm learns the noise from the data itself, without a priori knowledge of its sources. Firstly, we show how the proposed protocol can improve the averaged fidelity of reconstruction over linear inversion and maximum likelihood estimation in the finite-statistics regime, reducing at least by an order of magnitude the amount of necessary training data. 
Next, we demonstrate its use for out-of-distribution data in realistic scenarios. In particular, we consider squeezed states of few spins in the presence of depolarizing noise and measurement calibration errors and certify its metrologically useful entanglement content. The protocol introduced here targets experiments involving few degrees of freedom and afflicted by a significant amount of unspecified noise. These include NISQ devices and platforms such as trapped ions or photonic qudits.
\end{abstract}

\maketitle

\section{Introduction}
Modern quantum technologies are fueled by resources such as coherence, quantum entanglement, and Bell nonlocality. Thus, a necessary step to assess the advantage they may provide is the certification of the above features \cite{GUHNE20091, PhysRevLett.107.210404,PRXQuantum.2.010201,PhysRevLett.125.150503,Friis2019,Eisert2020,Sotnikov2022,9996689,PRXQuantum.3.010317,10.21468/SciPostPhysCore.6.2.028,Hangleiter_2017,plodzien2024entanglement}. The resource content of a preparation is revealed from the statistics (e.g. correlations) the device is able to generate. Within the quantum formalism, such data is encoded in the density matrix.
In a real-world scenario, the density matrix can be reconstructed based on finite information from experimentally available probes, a process known as quantum state tomography (QST) \cite{PhysRevLett.74.4101,PhysRevLett.83.3103,PhysRevLett.92.220402, Haffner2005,PhysRevLett.105.150401,5714248,Toth2010,Moroder_2012,Cramer2010,PhysRevLett.111.020401,Lanyon2017}. Hence, QST is a prerequisite for any verification task. On the other hand, the second quantum revolution brought new experimental techniques to control and manipulate quantum systems \cite{Acin_2018, Kinos2021,Laucht_2021,Zwiller2022,Fraxanet2022}, challenging established QST protocols. Both reasons elicited an extremely active field of research that over the years offered a plethora of algorithmic techniques \cite{Gross2010,Flammia_2012,Huszar2012,Kravtsov2013,Granade_2017,Toth2010,Schwemmer2014,Huang2020,Aaronson2020,PhysRevLett.124.010405,Cattaneo2023, Huang2024}.

Any conventional QST protocol suffers inevitably from a plethora of noise sources; ranging from measurement calibration errors, dark counts, losses, or technical noises, to name a few. Such effects are eminently challenging to a model and will eventually decohere the system and wash out any quantum resource. 

In recent years, machine learning (ML), artificial neural networks, and deep learning have entered the field of quantum technologies \cite{dawid2022modern}, 
offering many new solutions to QST \cite{Torlai2018,Carrasquilla2019,Cha2021,Melkani2020,Schmale2022,Xin2019, lidiak2022quantum,  Torlai2023, Akhtar2023scalableflexible, li2023efficient, kuzmin2023learning}.
These approaches learn the noise from the experimental data itself and are unaware of its sources or models ~\cite{adri1,Pan2022,koutny2022PRA,Shahnawaz2021,ma2023attentionbased}. Thus, not only shot noise, which is inherent to the QST task, but also other disturbances are susceptible to be mitigated by such methods. As only minimal assumptions about the system are required, they are especially suited for the certification task.

Despite their initial success, there are still setbacks in the application of neural network-based methods for QST. In this work, we will focus on two of them: (i) the learning ability of a network with a reduced training set size~\cite{Westerhout2020}, and (ii) the possibility of out-of-distribution (OOD) use of this class of methods \cite{OOD-Guerin2023}. OOD is a subfield of ML that analyzes how models perform on new data that do not belong to the training data distribution, with the latter called the \textit{in-distribution} dataset (ID). To this end, we offer a computationally fast general protocol that combines established QST protocols and a supervised architecture trained to denoise the density matrix.

To begin with, we assess the learning and generalization abilities of the proposed method with generic states, as produced e.g. from a random emitter. Then, we use our denoising network, trained exclusively on density matrices affected by shot noise only, to reconstruct new ones obtained from a simulated real-case scenario. In particular, we consider squeezed states of few spins under depolarizing and measurement calibration noise, and we certify its entanglement depth and usefulness for metrology applications.

%From the perspective of the ansatz under consideration, we note that, in addition, any \emph{a priori} assumption on the state, e.g. the pure-state ansatz~\cite{Torlai2019}, is inadmissible.

The new protocol aims to certify quantum resources in low-dimensional systems afflicted by a significant level of unknown noise \cite{Bent2015, Stricker2022,adri1}, where complete tomography is required. \\

The paper is organized as follows: in Sec.~\ref{sec:Preliminaries}, we introduce the main concepts behind the QST; in Sec.~\ref{sec:Methods} we introduce the data generation protocol and neural network architecture, as well as define QST as a denoising task. Sec.~\ref{sec:Results} is devoted to benchmarking our method against known approaches, and we test it on quantum states of physical interest. In Sec.~\ref{sec:concrete}, we provide practical instructions to implement our algorithm in an experimental setting. We conclude in Sec.~\ref{sec: conclusion} with several possible future research directions. 

\section{Preliminaries }\label{sec:Preliminaries}

Consider the $d$-dimensional Hilbert space. A set of informationally complete (IC) measurement operators $\hat{\boldsymbol{\pi}}=\{ \hat{\pi}_i \}$, $i = 1, \dots, d^2$, in principle, allows unequivocally reconstructing the underlying target quantum state $\hat{\tau} \in \mathbb{C}^{d \times d}$
within the limit of an infinite number of ideal measurements \cite{Fuchs2013, Fuchs2017}. 
After infinitely many measurements, one can infer the mean values
\begin{equation}
\label{eq:p_Born}
p_{i}  = \mathrm{Tr}[\hat{\tau} \hat{\pi}_{i}],
\end{equation}
and construct a valid vector of probabilities $\mathbf{p} = \{p_i \}$
for any proper state $\hat{\tau}\in \mathcal{S}$, where by $\mathcal{S}$ we denote the set of $d$-dimensional quantum states, i.e., containing all unit-trace, positive semi-definite (PSD) $d\times d$ Hermitian matrices. 
Alternatively, $\hat{\boldsymbol{\pi}}$ can form a set of operators that spans the space of Hermitian matrices. In such a case, $\mathbf{p}$ can be evaluated from multiple measurement settings (e.g., Pauli basis) and is generally no longer a probability distribution.  In any case, there exists a one-to-one mapping $Q$ from the mean values $\mathbf{p}$ to the target density matrix $\hat{\tau}$: 
\begin{align}
\label{eq:Q}
  Q\!:\,&\mathcal{F}_{\mathcal{S}}\longrightarrow \mathcal{S} \\
\nonumber
  &\mathbf{p}\longmapsto Q[\mathbf{p}]=\hat{\tau},
\end{align}
where $\mathcal{F}_{\mathcal{S}}$ is the space of accessible probability vectors. In particular, by inverting the Born's rule, Eq.~\eqref{eq:p_Born}, elementary linear algebra allows us to describe the map $Q$ as
\begin{equation}\label{eq:Q_p}
Q[\mathbf{p}] = \mathbf{p}^T \hat{G}^{-1}\hat{\boldsymbol{\pi}},
\end{equation}
where $\hat{G}$ is the Gram matrix of the measurements settings, with components $G_{ij} = \mathrm{Tr}(\hat{\pi}_i\hat{\pi}_j)$.  

The inference of the mean values $\mathbf{p}$ is only perfect in the limit of an infinite number of measurement shots, $N\rightarrow \infty$. 

In a realistic scenario, with a finite number of experimental runs $N$, we have access to frequencies of relative occurrence $\mathbf{f} = \{f_i := n_i/N\}$, where $n_i$ is the number of times the outcome $i$ is observed. 
Such counts allow us to estimate $\mathbf{p}$ within an unavoidable error dictated by the shot noise, whose amplitude typically scales as $1/\sqrt{N}$ \cite{QObook}. 
With only frequencies $\mathbf{f}$ available, we can use mapping $Q$ for estimation $\hat{\rho}$ of the target density matrix $\hat{\tau}$, i.e.,
\begin{equation}\label{eq:QST_f}
 \hat{\rho} = Q[\mathbf{f}].
\end{equation}
In the infinite number of trials $N\to \infty$, $f_i = p_i$ and $\hat{\rho} = \hat{\tau}$. Yet, in the finite statistics regime, as considered in this work, the application of the mapping as defined in Eq.~\eqref{eq:Q_p} to the frequency vector $\mathbf{f}$ will generally lead to nonphysical results (i.e.\ $\hat{\rho}$ not PSD). 
In such case, as an example of proper mapping $Q$ we can consider different methods for standard tomography tasks, such as linear inversion (LI), or maximum likelihood estimation (MLE), see Appendix \ref{sec:app_classical}. As operators $\boldsymbol{\hat{\pi}}$, we consider positive operator-valued measures (POVMs) and a more experimentally appealing Pauli basis (check Appendix~\ref{app:data_generation}).

\section{Methods} \label{sec:Methods}

\begin{figure}[t!]
    \centering
    \includegraphics[width = 0.48\textwidth, right]{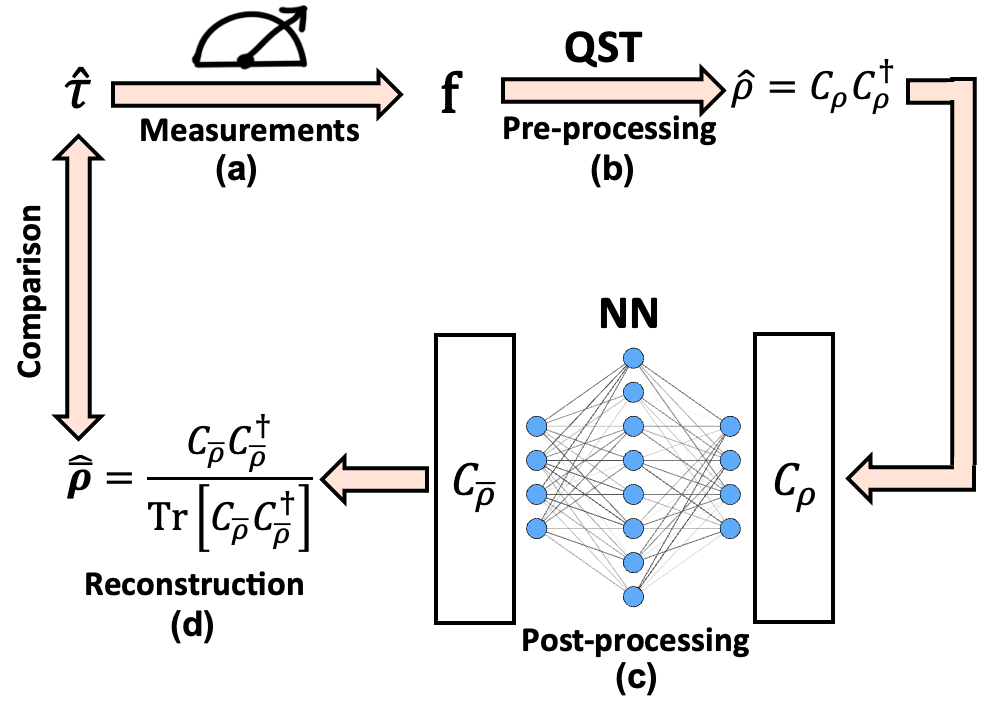}
    \caption{
    Schematic representation of the data pipeline of our QST hybrid protocol. Panel (a) shows data acquisition from a generic experimental set-up, during which the frequencies $\mathbf{f}$ are collected. Next, panel (b) presents standard density matrix reconstruction; in our work, we test the computationally cheap LI method together with the expensive MLE, to better analyze the network reconstruction behaviour and ability. Panel (c) depicts the matrix-to-matrix deep-learning strategy for Cholesky matrices reconstruction. The architecture herein considered combines convolutional layers for input and output and a transformer model in between. Finally, we compare the reconstructed state $\hat{\bar{\rho}}$ with the target $\hat{\tau}$.
    }
    \label{fig:sketch}
\end{figure}

This section describes our density matrix reconstruction protocol, data generation, neural network training, and inference procedure. In Fig.~\ref{fig:sketch}, we show how these elements interact within the data flow. In the following paragraphs, we elaborate on the proposed protocol in detail.

The first step in our density matrix reconstruction protocol, called \textit{pre-processing}, is a reconstruction of density matrix $\hat{\rho}$ using finite-statistic QST with frequencies $\mathbf{f}$ obtained from measurement prepared in target state $\hat{\tau}$. Next, we feed the reconstructed density matrix $\hat{\rho}$ through our neural network acting as a noise filter, which we call this stage \textit{post-processing}. To enforce the positivity of the neural network output, we employ the Cholesky decomposition of the density matrices, that is, $\hat{\rho} = C_\rho C_\rho^\dagger$ and $\hat{\tau} = C_\tau C_\tau^\dagger$, where $C_{\rho,\tau}$ are lower-triangular matrices. 
Such decomposition is uniquely provided that $\hat{\rho}$ and $\hat{\tau}$ are positive \footnote{If it is rank-deficient (e.g., for pure states), we add a small correction of amplitude order $10^{-5}$ to ensure state positivity.}.   
We treat the Cholesky matrix $C_\rho$ obtained from the finite-statistic QST protocol as a \textit{noisy} version of the target Cholesky matrix without noise $C_\tau$ calculated from $\hat{\tau}$. With these data, we prepare a supervised training for our architecture.

\subsection{Data generation}

To construct the training data set, we first start with generating $N_{\rm train}$ Haar-random $d$-dimensional target density matrices, $\{ \hat{\tau}_m \}$, where $m = 1,\dots, N_{\rm train}$. Next, we simulate experimental measurement outcomes $\mathbf{f}_m$, for each $\hat{\tau}_m$, in one of the two ways:
\begin{enumerate}
\item\textit{Directly}: When the measurement operators $\hat{\boldsymbol{\pi}}$ form an IC-POVM, we can take into account the noise by simply simulating the experiment and extracting the corresponding frequency vector $\mathbf{f}_m = \{n_i/N\}_{m}$, where $N$ is the total number of shots (i.i.d. trials) and the counts $\{n_i\}_{m}$ are sampled from the multinomial distribution.   
\item\textit{Indirectly}: As introduced in the preliminaries (Sec.~\ref{sec:Preliminaries}), with projective measurements $\hat{\boldsymbol{\pi}}$ the $\mathbf{p}_m$ is no longer a probability distribution, like the Pauli basis (see Appendix~\ref{app:data_generation}). So, we can insert an amount of noise as the \textit{direct} case, obtaining $\mathbf{f}_m = \mathbf{p}_m + \delta\mathbf{p}_m$, where $\delta\mathbf{p}_m$ is sampled from the multi-normal $\mathcal{N}\big(\mathbf{0}, \sim\mathbf{1}/(2\sqrt{N})\big)$ of mean zero and isotropic variance, saturating the shot noise limit.
\end{enumerate}

Upon preparing the frequency vectors $\{ \mathbf{f}_m\}$, we apply QST by mapping $Q$, Eq.~\eqref{eq:QST_f}, obtaining the set of reconstructed density matrices $\{ \hat{\rho}_m \}$. We employ a rudimentary and scalable method, i.e., linear inversion \footnote{After removal of negative eigenvalues}, but other QST methods can also be used. Finally, we construct the training dataset as $N_{\rm train}$ pairs $\big\{ \vec{C}_{\rho}, \vec{C}_{\tau} \big\}$, whereby $\vec{C}$ we indicate the vectorization (flattening) of the Cholesky matrix $C$ (see Appendix~\ref{mse justification} for definition).

\subsection{Neural network architecture}

Our proposed architecture is inspired by other recent models \cite{Zhu2022,Requena2022,ma2023attentionbased}, combining convolutional layers with a transformer layer that implements a self-attention mechanism \cite{bahdanau2016neural,vaswani2023attention}. The convolutional layer extracts local features from the data, while the transformer seizes global ones. By combining them, we aim at taking the best of both approaches. The self-attention mechanism utilizes the representation of the input data as nodes within a graph \cite{dwivedi2021generalization} and aggregates the relationships between the nodes. \\

\noindent \textbf{Architecture}.-- The neural network action can be described as a mapping $h_{\boldsymbol{\theta}}$ that transforms the input - vectorized Cholesky matrix $\hat{C}_{\rho}$ into an output $h_{\boldsymbol{\theta}}(\hat{C}_{\rho})$. The symbol $\boldsymbol{\theta}$ denotes all the variational parameters as weights and biases to be optimized during the training phase.   
The choice of architecture considered here contains two convolutional layers $h_\mathrm{cnn}$, a transformer layer $h_\mathrm{tr}$ between, and a final linear layer $h_\mathrm{l}$,  i.e.:
\begin{align}
\label{eq: transformer function}
     h_{\boldsymbol{\theta}}[\vec{C}_{\rho}] = {\rm tanh} 
     (h_{\mathrm{\mathrm{l}}})\circ h_{\mathrm{tr}} \circ  \gamma(h_{\rm cnn})[\vec{C}_{\rho}] \ ,
\end{align}
 where $\gamma(y) = 1/2y(1 + \text{Erf}(y)/\sqrt{2})$, $y\in\mathbb{R}$, is the Gaussian Error Linear Unit (GELU) activation function \cite{GELU}, broadly used in the modern transformers architectures,
and tanh(y) is the hyperbolic tangent, acting element-wise on neural network nodes. A detailed explication of the model is offered in App.~\ref{app: architecture details}.

\subsection{Neural network training}
 
%\textcolor{teal}{To design a denoising task, we train our neural network $h_{\boldsymbol{\theta}}$ in a supervised way, preparing matrix-to-matrix mapping in its vectorized form $h_{\boldsymbol{\theta}}: \vec{C}_{\rho}\to \vec{C}_{\boldsymbol{\theta}}$.} 
The neural network training process relies on minimizing the cost function defined as a mean squared error (MSE) of the network output with respect to the target density matrix $\hat{\tau}$.

\begin{equation}
\label{eq:costf}
    {\cal L}(\boldsymbol{\theta})  = \| \vec{C}_{\tau}-\vec{C}_{\boldsymbol{\theta}}\|^2
    +{\rm Tr}[C_{\boldsymbol{\theta}}C_{\boldsymbol{\theta}}^\dagger],
\end{equation}
with ${\rm Tr}[C_{\boldsymbol{\theta}}C_{\boldsymbol{\theta}}^\dagger]$ a regularization term, cf.\ chapter 7 of Ref.~\cite{Goodfellow-et-al-2016} for detail. 
We train the model with a dataset containing $N_{\rm train}$ training samples $\{\hat{\rho}_l \}$. The equivalence between MSE and the Hilbert-Schmidt (HS) distance is discussed in detail in Appendix~\ref{mse justification}, where we also demonstrate that the mean squared error used in the cost function, Eq.~\eqref{eq:costf}, is a natural upper bound of the quantum fidelity. Hence, the choice of the cost function, Eq.~\eqref{eq:costf} approximates the target state in a proper quantum metric. . 

By minimizing Eq.~\eqref{eq:costf}, we obtain the set of optimal parameters $\bar{\boldsymbol{\theta}} $ for our model
$h_{\bar{\boldsymbol{\theta}}}$.
Finally, the neural network allows for the reconstruction of the target density matrix $\hat{\tau}$ via Cholesky matrix $C_{\bar{\rho}}$ \footnote{Notice that $C_{\bar{\rho}}$ is not necessarily a Cholesky matrix in its canonical form (e.g., with diagonal elements positive), as such constraints are not imposed. However, this fact does not pose further issues as in any case, by construction, post-processing will lead to a proper state $\hat{\bar{\rho}}$.}, i.e.,
\begin{equation}\label{eq:inference}
    \hat{\bar{\rho}} = \frac{C_{\bar{\rho}} C_{\bar{\rho}} ^\dagger}{\Tr[C_{\bar{\rho}} C_{\bar{\rho}} ^\dagger]} \simeq \hat{\tau},
\end{equation}
where $C_{\bar{\rho}}$ is reshaped from $\vec{C}_{\bar{\rho}} =h_{\bar{\boldsymbol{\theta}}}[\vec{C}_\rho]$.
\newline

\section{Results and discussion}
\label{sec:Results}

Following the presentation of our QST protocol, we demonstrate its advantages in scenarios of both computational and physical interest. To this aim, we consider two examples.

As the first example, we study an idealized random quantum emitter (see e.g. Refs.~\cite{Liu2023,Tang2022} for recent experimental proposals) that samples high-dimensional mixed states from the Hilbert-Schmidt distribution. After probing the system using single-setting square-root POVM, we can show the usefulness of our neural network by improving the  preprocessed LI and MLE states. This example enables us to assess the learning ability and expressivity of the QST neural network introduced in this work.

In the second example, we focus on a specific class of muti-qubit pure states of special physical relevance, i.e., with metrological resource, as quantified by the quantum Fisher information (QFI). Such states are generated via one-axis twisting dynamics (OAT) \cite{Kitagawa1993,wineland1994squeezed}.
 We simulate realistic data by considering the experiment in the presence of depolarizing noise, which mixes the OAT state $\ket{\Psi}$ according to 
\begin{equation}
\label{depo channel}
    \hat{\sigma} = (1-p)\ketbra{\Psi} + \frac{p\mathbb{I}}{d},
\end{equation}
where $\mathbb{I}$ is the identity operator. 
Furthermore, we reproduce miscalibration of the measurements by adding a bias error to the inferred expectation values. This example allows us to evaluate our protocol with OOD data.  

%Finally, in Appendix~\ref{OAT ideal} we perform another OOD analysis of our protocol considering only sampling noise, but comparing the reconstructions obtained from Pauli operators with the SIC-POVM, which are an optimal choice of measurement operators for tomographic tasks.}

\subsection{Reconstructing high-dimensional random quantum states}
\label{app: mixed states}
\noindent\textbf{Scenario}.-- Let us consider states $\{\hat{\tau}_j\}$ sampled from distribution uniform with respect to the Hilbert-Schmidt measure (see Appendix~\ref{app:random}) on the Hilbert space 
of dimension $d=9$.
Each target state $\hat{\tau}_j$ is measured $N_{\rm trial}$ times using its copies. 
This scenario allows us to benchmark our algorithm against the protocol offered in Ref.~\cite{koutny2022PRA}.

We prepare measurements on each trial state $\hat{\tau}_j$ using information-complete square-root POVM (IC), as defined in Eq.~\eqref{eq: root square povm}.
This allows obtaining the state reconstruction $\hat{\bar{\rho}}_j$ via two standard QST protocols, i.e. linear inversion (LI) and maximum likelihood estimation (MLE) algorithms, as well as by our neural network enhanced protocols denoted LI-NN and MLE-NN, see Fig.~\ref{fig:sketch}. Finally, we evaluate the quality of the reconstruction using the square of the Hilbert-Schmidt distance between the target and the reconstructed state $ D_{\rm HS}^2(\hat{\bar{\rho}}_j,\hat{\tau}_j) = \mathrm{Tr}[(\hat{\bar{\rho}}_j-\hat{\tau}_j)^2]$. \\
 
 \begin{figure}[ht!]
    \includegraphics[width=\linewidth]{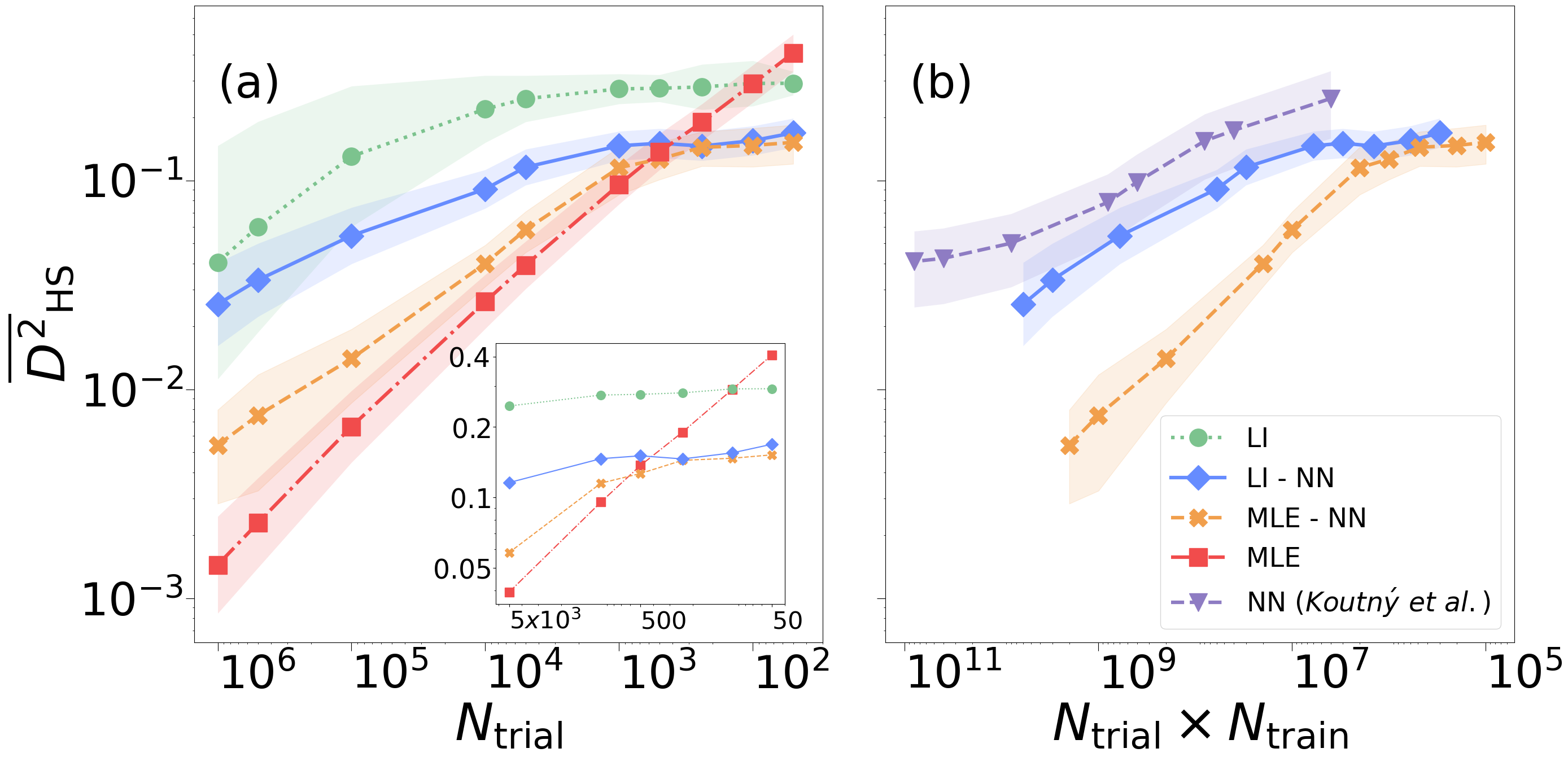}
   \caption{
   Evaluation of the QST reconstruction quality measured by the mean value of the Hilbert-Schmidt distance square, $\overline{D^2}_{\rm HS}$, between the target and the reconstructed state for different QST protocols, averaged over $1000$ target states.
   In both panels, the best-performing setups are those that are as far right (better quality) and bottom (less costly) as possible. 
   Panel (a) uses the number of measurements $N_{\rm trial}$ to compare four QST protocols: linear inversion (LI, green dots, neural network enhanced MLE (MLE-NN, orange crosses), neural network enhanced LI (NN-LI, blue diamonds) and maximal likelihood estimation (MLE, red squares). We add an inset focusing in the undersampled regime, $N_{\rm trial}\leq 5\times 10^3$.
   Panel (b) shows the quality of reconstruction as a function of product $N_{\rm trial}\times N_{\rm train}$ for the latter two protocols and the network model proposed in Ref.~\cite{koutny2022PRA} (violet triangles). 
   Both panels depict resource costs on horizontal axes in different scenarios: in (a), the cost is the number of performed measurements, while in (b), the training phase is additionally counted as a cost. 
   Our proposed protocol achieves competitive averaged HS reconstruction for the size of training data of an order of magnitude smaller than the method proposed in Ref.~\cite{koutny2022PRA}. 
   During models' training, we used $N_{\rm train} = 2000$ random pure states for the MLE-NN protocol, and $N_{\rm train}= 5000$ for the LI-NN.
   Lines are to guide the eye; shadow areas represent one standard deviation.}
   \label{fig: benchmark}
 \end{figure}

\noindent\textbf{Benchmarking}.--
Fig.~\ref{fig: benchmark}(a) presents the averaged Hilbert-Schmidt distance square $\overline{D^2}_{\rm HS}$ as a function of the number of trials states $N_{\rm trial}$.
To obtain a reconstruction for a given averaged HS distance, our neural network-enhanced protocols require a much lower number of $N_{\rm trial}$ copies of states, compared to linear inversion and maximum likelihood estimation alone.
We note that the proposed protocols improve MLE for the relatively small number of trial states $N_{\rm trial}<10^3$ (see inset), which is important from an experimental point of view.
As expected, the lowest HS distance is obtained for many trials of the MLE algorithm.

Recently, a state-of-the-art QST neural network protocol was proposed by Koutný et al.~\cite{koutny2022PRA}. The authors report better performance than the MLE and LI algorithms with $N_{\rm train} = 8\cdot 10^5$ training samples for qutrit as well as larger systems, $d \geq 3$.
In Fig.~\ref{fig: benchmark}(b), we show that our protocol requires an order of magnitude smaller training data, achieving a comparable level of reconstruction.

\subsection{Certifying metrologically useful entanglement in noisy spin-squeezed states}

In this simulation, we reconstruct a class of physically relevant multiqubit pure states in a realistic scenario. Specifically, we consider a chain of $L = 4$ spins-$1/2$ (Hilbert space of dimension $d=16$). The target quantum states are dynamically generated during the one-axis twisting (OAT) protocol \cite{Kitagawa1993,wineland1994squeezed} 
\begin{equation}
    \label{eq:def_OAT}
    \ket{\Psi(t)} = e^{-it\hat{J}_z^2}\ket{+}^{\otimes L} \ ,
\end{equation} 
where $\hat{J}_z$ is the collective spin operator along $z$-axis
and $\ket{+}^{\otimes L} = [(\ket{\uparrow} + \ket{\downarrow})/\sqrt{2}]^{\otimes L}$ is the initial state prepared in a coherent spin state along $x$-axis (orthogonal to $z$).
The OAT protocol generates spin-squeezed states useful for high-precision metrology, allowing to overcome the shot-noise limit~\cite{RevModPhys.90.035005,PhysRevLett.105.053601,wineland1994squeezed,MllerRigat2023}, as well as many-body entangled
and the many-body Bell correlated states~\cite{Tura1256,schmied2016bell,Aloy2019,Baccari2019,Tura2019,zukowski2002bell,cavalcanti2007bell,he2011entanglement,cavalcanti2011unified,spiny.milosz,PhysRevLett.126.210506,10.21468/SciPostPhysCore.5.2.025, MllerRigat2021}. 
OAT states have been extensively studied theoretically~\cite {PhysRevA.46.R6797,PhysRevA.47.5138,PhysRevA.92.043622,PhysRevLett.100.210401,Li2009,PhysRevA.96.013823,Kajtoch2018, Schulte2020,PhysRevLett.126.160402, PhysRevA.105.022625}, and can be realized with a variety of ultra-cold systems, utilizing atom-atom collisions~\cite{Treutlein2010,Oberthaler2010,Chapman2012,PhysRevLett.125.033401}, and atom-light interactions~\cite{PhysRevLett.104.073602,PhysRevLett.105.080403}. The recent theoretical proposals for the OAT simulation with ultra-cold atoms in optical lattices effectively simulate Hubbard and Heisenberg models~\cite{Kajtoch2018, PhysRevResearch.1.033075, Plodzien2020, Plodzien2022, Plodzie2023generation, PhysRevResearch.3.013178,Hernandez2022,Dziurawiec2023,yanes2023spin}.\\

\noindent\textbf{Data preparation}.-- For this task, we generate our data focusing on the experimentally friendly Pauli operators. The state under consideration are prepared \textit{indirectly}, with a fidelity from the target OAT state of $\sim85\%$. Then we simulate the presence of a noise channel by depolarizing our input state with a strength factor $p=0.3$ according to Eq.~\eqref{depo channel}. Lastly, to simulate the presence of a calibration defect in the measurement apparatus, a fixed bias of random values of the order of $10^{-4}$ is added to the Born values. After applying all these noise sources, the LI reconstructions can obtain an average fidelity of $75.4\pm 1.1 \%$. 
%We start by training on a dataset with $N_{\rm train}=10,\, 500$ Haar random pure states, see Appendix~\ref{app:random}.
For the test set, we select 100 OAT states in evenly spaced times $t\in (0, \pi)$ and assess the average reconstruction achieved by our neural network \footnote{Code to generate the data and reproduce these plots are available in the github} .
%In the perspective of minimal amounts of experimental resources, in simulations, we measure each evolved state only once {\color{red} I don't understand what you mean by measuring only once the state}. 

The description of the model under consideration can be found in Table~\ref{tab:nn only sampling}.  \\

%We compare the values with the average score obtained for generic Haar-distributed states, which is the set on which the model was trained. 

\begin{figure}[t!]
    \centering
    \includegraphics[width = \linewidth]{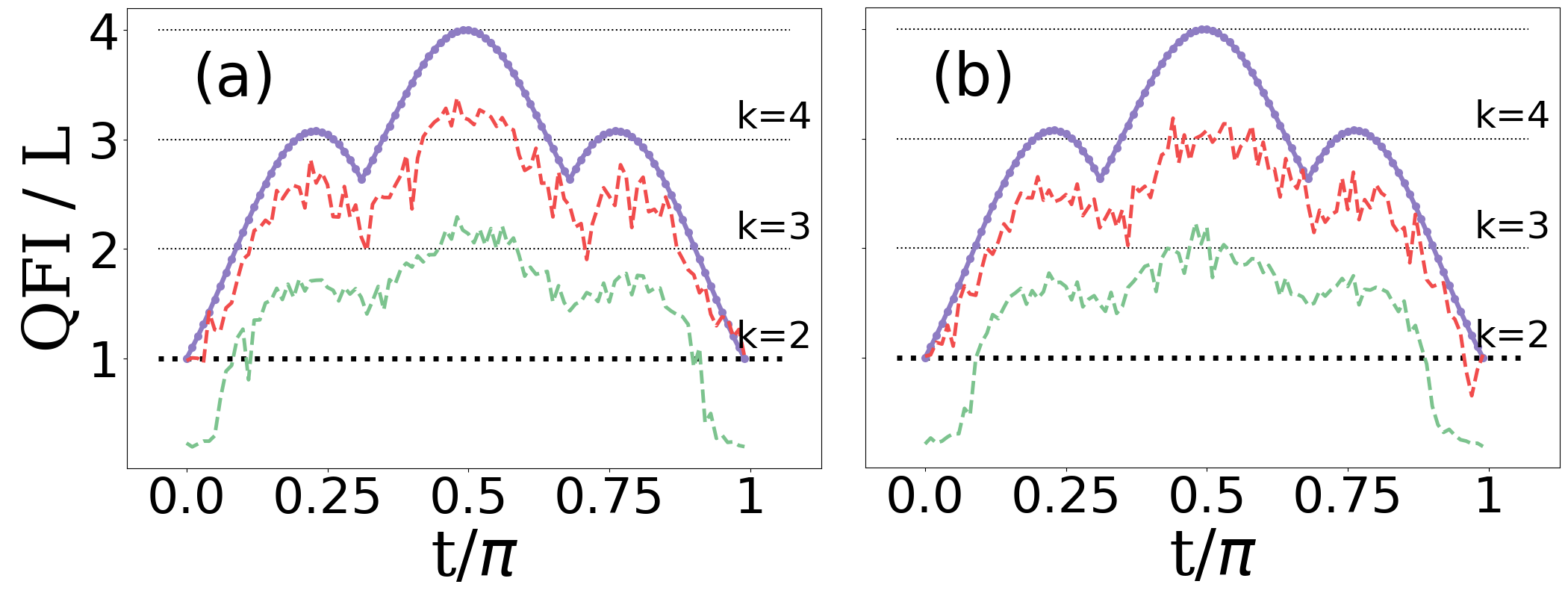}
    \caption{Two different simulations for out-of-distribution (OOD) inference. In each panel, we evaluate the normalized Quantum Fisher Information (QFI) for 100 four-qubit states as validation metric. The target, noiseless states are evolved according to the OAT dynamics given in \eqref{eq:def_OAT}, and depicted by the purple dotted line. For this OOD tests, the neural network was trained exclusively to learn statistical sampling noise. During inference, test data are permeated by depolarization and measurement (calibration) errors also. The green line represents the normalized QFI derived from reconstructions via the Linear Inversion (LI) algorithm; the red line illustrates the enhancement provided by the network when supplemented with LI algorithm reconstructions, underscoring the robustness of our protocol in mitigating noise effects.  }
    \label{fig:realistic OAT scenario}
\end{figure}

\noindent\textbf{Inferring the quantum Fisher information}.-- Finally, we evaluate the metrological usefulness of the reconstructed states measured by the quantum Fisher information (QFI), $F_{Q}[\hat{\rho},\hat{G}]$. The QFI is a non-linear function of the state and quantifies the sensitivity upon rotations generated by $\hat{G}$.
Notwithstanding, its usefulness in metrology, it is a highly nontrivial task to evaluate it experimentally due to its great sensitivity to noise disturbances, with state-of-the-art research only up to 4 qubits~\cite{Yu_2021,Yu_2022}.
We refer the reader to Appendix~\ref{sec: app_QFI} for more details. 

In this context, we consider the collective spin component $\hat{J}_{\mathbf{v}}$ as the generator $\hat{G} = \hat{J}_{\mathbf{v}}$, with the orientation $\boldsymbol{v}\in \mathbb{R}^3$ selected to achieve maximal sensitivity. Quantum Fisher Information (QFI) related to collective rotations can also serve to verify quantum entanglement \cite{MllerRigat2023}, specifically the entanglement depth $k$, which is the smallest number of genuinely entangled particles required to describe the state. If $F_Q[\hat{\rho}, \hat{J}_{\mathbf{v}}] > kL$, then the quantum state $\hat{\rho}$ possesses an entanglement depth of at least $k+1$ \cite{hyllus2012fisher,toth2012multipartite}. In particular, for states with depth $k = 1$ (i.e., separable), the absence of detected entanglement implies that the metrological capability is limited to the shot-noise threshold~\cite{pezze2009entanglement}. This limit is reached by coherent spin states, such as our initial ($t=0$) state for the evolution of the One-Axis Twisting (OAT), $\ket{+}^{\otimes L}$.

\vspace{3pt}
\noindent{\bf OOD Results.--} In Fig.~\ref{fig:realistic OAT scenario}, we present the evolution of the QFI (normalized by the coherent limit, $L = 4$) for the OAT target states (top of the solid blue lines). For this numerical experiment, we make full use of the OOD approach; using a network trained only for tackling sampling noise, we feed it in inference with a dataset that also considers the depolarization and measurement noise. As Fig.~\ref{fig:realistic OAT scenario} shows, for two different realizations of calibration noise, our network can highly improve the LI reconstructions obtaining a fidelity of $88.7\pm 2.3 \%$ and $91.0\pm 1.9 \%$, on the left and right panel, respectively. Thereby, we surpass the three-body bound ($\mathrm{QFI}/L = 3$), thus revealing a genuine 4-body entanglement, which is the highest depth possible in this system (since it is of size $L =4$). For example, note that at time $t =\pi/2$, the OAT dynamics generates the cat state, $|\Psi(t = \pi/2)\rangle = (e^{-i\pi/4}|+\rangle^{\otimes 4}+ e^{+i\pi/4}|-\rangle^{\otimes 4})/\sqrt{2} $, which is genuinely $L$-body entangled, and so it is certified. 
For completeness, two complexity analyses are offered. First, in Appendix~\ref{OAT ideal}, an analysis of the  QFI time evolution for OAT states for different sampling noise values only, using Pauli and tomographically optimal SIC-POVM operators in data generation. 
Next, in Appendix~\ref{app:benchmarking}, a benchmark of our model is shown against two different convolutional architectures to assess the advantage offered by our transformer-based model on quantum data. Lastly, we propose an alternative method to incorporate the statistical noise as a depolarizing channel in the Appendix~\ref{app:MLE_and_HS_metric}.

\section{Concrete experimental implementation}
\label{sec:concrete}

To recapitulate this contribution, as a complement to Fig.~\ref{fig:sketch} and our repository provided in Ref.~\cite{OurRepo}, we summarize the practical implementation of the protocol introduced in this work. 

\begin{enumerate}
    \item \textit{Scenario}: We consider a finite-dimensional quantum system prepared in a target state $\hat{\tau}$. Here, our objective is to verify the preparation of a quantum state $\hat{\tau}$   via QST. To this end, we set a particular measurement basis $\hat{\boldsymbol{\pi}}$ to probe the system.  
    \item \textit{Experiment}:  After a finite number of experimental runs, we construct the frequency vector $\mathbf{f}$ from the countings. 
    \item \textit{Preprocessed quantum state tomography}: From the frequency vector $\mathbf{f}$ and the basis $\hat{\boldsymbol{\pi}}$, we infer the first approximation of the state $\hat{\rho}$ via the desired QST protocol (e.g., one of those introduced in Appendix~\ref{sec:app_classical}).
    \item \textit{Assessing pre-reconstruction}: We evaluate the quality of the reconstruction by, for example, computing $D^2_{\mathrm{HS}}(\hat{\tau},\hat{\rho})$, quantum fidelity, or any other meaningful quantum metric. To improve such a score, we resort to our neural network solution to complete a denoising task. As with any deep-learning method, training is required.   
    \item \textit{Training strategies}: Different training strategies can be implemented:
    \begin{enumerate}
        \item Train over uniform ensembles (e.g., Haar, HS, Bures etc.) if $\hat{\tau}$ is a typical state or we do not have information about it. 
        \item Train over a subspace of states of interest. For example, if we reconstruct OAT states (Section~\ref{sec:Results}b), we may train only in the permutation-invariant sector. 
        \item Due to the greater generalization ability demonstrated when the case of mixed state was considered, we can perform transfer learning to tailor a pre-trained model on a specific apparatus with less computational resources; in this way, the model will have knowledge of that specific apparatus noise. For example, if we have a quantum random source to characterize (Section~\ref{sec:Results}a), an amount of experimental data corresponding to the $10-15\%$ of the training dataset can be used to refine the model. 
 
    \end{enumerate}
    \item \textit{Feeding the neural network}: We feed the preprocessed state $\hat{\rho}$ into our trained matrix-to-matrix neural network to recover the enhanced quantum state $\hat{\bar{\rho}}$.
    \item \textit{Assessing the neural network}: We compute the updated reconstruction metric on the post-processed state $D^2_{\mathrm{HS}}(\hat{\tau},\hat{\bar{\rho}})$. Finally, we assess the usefulness of the neural network by comparing how small such a value is compared to the pre-processed score $D^2_{\mathrm{HS}}(\hat{\tau},\hat{\rho})$. 
    
\end{enumerate}

The strength of our proposed protocol lies in its broad applicability, as the choice of the basis $\hat{\boldsymbol{\pi}}$ and the QST pre-processing method is arbitrary.

\section{Conclusions}\label{sec: conclusion}

We proposed a novel deep learning protocol that improves standard quantum state tomography methods, such as Linear Inversion and Maximum Likelihood Estimation. Based on a combination of transformer and convolutional layers, we greatly reduce the dataset dimension for training, and we can perform denoising on new unseen data, when also depolarization and measurement noise are accounted for. First, the proposed method reduces the number of necessary measurements in the target density matrix by at least an order of magnitude compared to other QST protocols supported by finite-statistic neural networks.

Secondly, for 4-qubits and Bell-correlated few-spin states generated with OAT, the inference stage was on OOD. We tested our model, pre-trained only for statistical sampling noise, on data accounting also for depolarization and measurement noise, achieving an average fidelity reconstruction $ 88.7\pm 2.3 \%$ and $91.0\pm 1.9\%$ for two different realizations of noise in the measurement setup. The superior learning ability demonstrated for mixed states makes our architecture an optimal candidate for transfer learning, when further refinement is desired on a pre-trained model using fewer experimental data. On the other hand, the OOD demonstrate the potential of our protocol for more plug-and-measure applications, given its high resilience to unknown level of noise due to different physical sources. Thus, it paves the way for the use of these novel methods in current quantum computers, NISQ devices, and quantum simulators based on spin arrays \cite{OszmaniecBosonicStates, Toth2010}.

Another persistent challenge in this field concerns the scalability of the algorithms with the number of subsystems. In this regard, several strategies are proposed based on \textit{incomplete} tomography like the well known generative NN applications~\cite{Carrasquilla2019}. A limitation of the protocol is its scalability, which is limited by the use of Cholesky decomposition in data preprocessing, which limits our method to applications of modest Hilbert space dimensions. An extension to a Cholesky-free approach is left for future explorations.   \\

\noindent\textbf{Data and code availability}.-- Data and code are available at Ref.~\cite{OurRepo}. \\

\section*{ACKNOWLEDGMENTS}

The authors thank Leonardo Zambrano, Federico Bianchi, and Emilia Witkowska for the fruitful discussions.
% ICFO
ICFO group acknowledges support from:  Europea Research Council AdG NOQIA; MCIN/AEI (PGC2018-0910.13039/501100011033, CEX2019-000910-S/10.13039/501100011033, Plan National FIDEUA PID2019-106901GB-I00, Plan National STAMEENA PID2022-139099NB, I00, project funded by MCIN/AEI/10.13039/501100011033 and by the “European Union NextGenerationEU/PRTR" (PRTR-C17.I1), FPI); QUANTERA MAQS PCI2019-111828-2); QUANTERA DYNAMITE PCI2022-132919, QuantERA II Programme co-funded by European Union’s Horizon 2020 program under Grant Agreement No 101017733); Ministry for Digital Transformation and of Civil Service of the Spanish Government through the QUANTUM ENIA project call - Quantum Spain project, and by the European Union through the Recovery, Transformation and Resilience Plan - NextGenerationEU within the framework of the Digital Spain 2026 Agenda; Fundació Cellex; Fundació Mir-Puig; Generalitat de Catalunya (European Social Fund FEDER and CERCA program, AGAUR Grant No. 2021 SGR 01452, QuantumCAT \ U16-011424, co-funded by ERDF Operational Program of Catalonia 2014-2020); Barcelona Supercomputing Center MareNostrum (FI-2023-3-0024); Funded by the European Union. Views and opinions expressed are however those of the author(s) only and do not necessarily reflect those of the European Union, European Commission, European Climate, Infrastructure and Environment Executive Agency (CINEA), or any other granting authority.  Neither the European Union nor any granting authority can be held responsible for them (HORIZON-CL4-2022-QUANTUM-02-SGA  PASQuanS2.1, 101113690, EU Horizon 2020 FET-OPEN OPTOlogic, Grant No 899794),  EU Horizon Europe Program (This project has received funding from the European Union’s Horizon Europe research and innovation program under grant agreement No 101080086 NeQSTGrant Agreement 101080086 — NeQST); ICFO Internal “QuantumGaudi” project; European Union’s Horizon 2020 program under the Marie Sklodowska-Curie grant agreement No 847648;“La Caixa” Junior Leaders fellowships, La Caixa” Foundation (ID 100010434): CF/BQ/PR23/11980043.
% AKS:
AKS acknowledges support from the European Union’s Horizon 2020 Research and Innovation Programme under the Marie Skłodowska-Curie Grant Agreement No. 847517.
% MP:
M.P. acknowledges the support of the Polish National Agency for Academic Exchange, the Bekker programme no:
PPN/BEK/2020/1/00317.
Views and opinions expressed are, however, those of the authors only and do not necessarily reflect those of the European Union, European Commission, European Climate, Infrastructure and Environment Executive Agency (CINEA), nor any other granting authority. Neither the European Union nor any granting authority can be held responsible for them.

\bibliography{bibliography.bib}

\appendix

\section{Established approaches for state tomography}
\label{sec:app_classical}

We review four of the most well-known approaches to quantum state reconstruction that could be potentially improved with the novel protocol proposed in this work. \\

\noindent\textbf{Linear inversion (LI)}.-- By inverting Born's rule Eq.~\eqref{eq:p_Born} (main text) we can express the state dependence on the mean values $\mathbf{p} = \{p_i \}$. 
\begin{equation}
    \label{eq:LI}
    \hat{\tau} = \mathbf{p}^T \mathrm{Tr}[\hat{\boldsymbol{\pi}}\hat{\boldsymbol{\pi}}^T]^{-1} \hat{\boldsymbol{\pi}} \ .
\end{equation}
Note that the inverse of the Gram matrix $\mathrm{Tr}(\hat{\boldsymbol{\pi}}\hat{\boldsymbol{\pi}}^T)$ exists as the basis is informationally-complete (IC). If it is (informationally) overcomplete, one needs to replace the inverse with the pseudoinverse. Finally, if it is under-complete (only partial information is available), it will determine the state up to a linear subspace.  

The LI method infers $\hat{\rho}$ by replacing in Eq.~\eqref{eq:LI} the ideal expectation values $\mathbf{p}$ with the vector $\mathbf{f}$ of the experimental frequencies (counts). This naive substitution generally leads to a negative matrix, $\hat{\rho}\nsucceq 0$. An \textit{optimal} way to tame its negative eigenvalues was given in~\cite{Smolin2012} by finding the physical state closest to $\hat{\rho}$ in the 2-norm. The drawback of LI is the fact that it can be affected by any type of noise.\\

\noindent\textbf{Least-squares estimation (LSE)}.-- Here, the reconstructed state is chosen to minimize the mean square error between the experimental frequencies $\mathbf{f}$ and the state probability distribution $\mathrm{Tr}(\hat{\boldsymbol{\pi}}\hat{\rho})$. The resulting problem can be expressed as,
\begin{align}
\label{eq:LSE_problem}
  \hat{\rho}_{\mathrm{LSE}} =  \arg\min_{\hat{\rho}\succeq 0} |\mathbf{f}- \mathrm{Tr}(\hat{\boldsymbol{\pi}}\hat{\rho})|^2 .
\end{align}
The problem Eq.~\eqref{eq:LSE_problem} is convex and, in addition, it is a disciplined convex problem (DCP), i.e., a class of convex optimizations that are efficiently addressed with commercially available solvers such as MOSEK \cite{mosek}.  \\

\noindent\textbf{Maximum likelihood estimation (MLE)}.-- In this case, the reconstructed state maximizes the likelihood of having produced the observed experimental outcomes, 
\begin{equation}
\hat{\rho}_{\mathrm{MLE}}=\arg\max_{\hat{\rho}\succeq 0} \log P(\hat{\rho}|\mathbf{f}).
\end{equation}
Our counting experiment is modeled as a multinomial. Consequently, the log-likelihood is $\log P(\hat{\rho}|\mathbf{f}) = \mathbf{f}\cdot \log(\mathrm{Tr}[\hat{\boldsymbol{\pi}}\hat{\rho})])$, which is a concave function of the state, but the resulting task is not a DCP. 
Therefore, solving it can be expensive, especially for a large Hilbert dimension space $d$.   
The MLE is a robust estimator against noise; however, it is computationally demanding, suffering from the exponential scaling of the inputs.   \\

\section{Informationally complete measurement operators}
\label{app:data_generation}

To obtain our set of initial Born values, we use three informationally complete (IC) sets of Hermitian operators $\boldsymbol{\pi}$ defined on the Hilbert space of $d$. \\

\noindent\textbf{Square-root POVM}.-- The first set $\boldsymbol{\pi}$ of measurement operators consists of POVM generated by the square root measurements, defined as
\begin{equation}
    \pi_i = \hat{H}^{-1/2}\ket{\phi_i}\bra{\phi_i}\hat{H}^{-1/2} \ \quad \hat{H} = \sum_{i\in [d^2]} \ket{\phi_i}\bra{\phi_i},
    \label{eq: root square povm}
\end{equation}
with $\{\ket{\phi_i}\in \mathcal{H}\}_{i\in [d^2]}$ are randomly generated Haar distributed pure states (see Appendix~\ref{app:random}). \\

\noindent\textbf{SIC-POVM}.-- The second set $\boldsymbol{\pi}$ of measurement basis operates in an $L$-qubit system. The basis consists of the tensor product of local sic-POVM, constructed by using the local vectors:
\begin{equation}\label{eq:POVMtetrahedral}
    \begin{split}
  \mathbf{s}_1 &= (0,0,1) \\
  \mathbf{s}_2 &=(\tfrac{2\sqrt{2}}{3}, 0, -\tfrac{1}{3} )\\
  \mathbf{s}_3 &=( -\tfrac{\sqrt{2}}{3},\sqrt{\tfrac{2}{3}},-\tfrac{1}{3} )\\
  \mathbf{s}_4 &=  ( 
 -\tfrac{\sqrt{2}}{3},-\sqrt{\tfrac{2}{3}},-\tfrac{1}{3} )  . 
    \end{split}
\end{equation}
The space of Hermitian operators acting in the global Hilbert space $\mathcal{H} = [\mathbb{C}^2]^{\otimes L}$ can then be spanned by
\begin{equation}
    \hat{\boldsymbol{\pi}} = \left\{\hat{\pi}_{(a,b)\cong i}=\bigotimes_{b\in [L]}\frac{\hat{\sigma}_0 + \mathbf{s}_{a_b}\cdot \hat{\boldsymbol{\sigma}}}{4}  \right\}_{\{a_b \}_{b\in [L]}\in [4]^L} \ ,
\end{equation}
where $\hat{\boldsymbol{\sigma}}=(\hat{\sigma}_x, \hat{\sigma}_y, \hat{\sigma}_z)$ is the Pauli vector and $\hat{\sigma}_0 = \mathbb{I}_2$ is the identity acting in the local space $\mathbb{C}^2$. 

Note that for any properly normalized state $\hat{\tau}$, $\mathbf{p}:=\mathrm{Tr}(\hat{\boldsymbol{\pi}}\hat{\tau})$ constitutes a valid probabillity distribution
\begin{equation}
    \forall i \in \left[2^{2L}\right],\ \  p_i\geq 0 \mbox{ and } \sum_{i\in \left[2^{2L}\right]}p_i = 1
\end{equation}
This observation is also true for the previous basis (square-root POVM) that we reviewed. \\

\noindent\textbf{Pauli basis}.-- The last IC basis $\hat{\boldsymbol{\pi}}$ in $L$-qubit systems is the Pauli basis constructed as: 
\begin{equation}
    \hat{\boldsymbol{\pi}} = \left\{\bigotimes_{b\in [L]} \hat{\sigma}_{a_b}  \right\}_{\{a_b \}_{b\in [L]}\in \{0,x,y,z \}^L} \ . 
\end{equation}
With respect to such a basis, expectation values can be evaluated experimentally by rotation of each qubit individually. This is also true for the SIC-POVM if evaluated with multiple settings. Such expectation values $\mathbf{p}$ no longer form a probability distribution (note that, in particular, such mean values can be negative). The reason why it will not lead to a probability distribution is that the basis does not form a POVM (that is, its elements are not PSD and do not sum up to $\mathbb{I}$). However, it covers the whole space of Hermitian matrices supported in $[\mathbb{C}^{2}]^{\otimes L}= \mathcal{H}$ as any basis specified in this appendix.

\section{From quantum fidelity to mean-squared error}
\label{mse justification}

\noindent\textbf{Upper bound on the Bures distance}.-- The Bures distance between two states $\hat{\rho}$ and $\hat{\tau}$ is defined as,
\begin{equation}
    D_B(\hat{\rho},\hat{\tau}) = 2 - 2\sqrt{F(\hat{\rho}, \hat{\tau})},
\end{equation}
where,
\begin{equation}
\sqrt{F(\hat{\rho}, \hat{\tau})} = \mathrm{Tr}\left[\sqrt{\sqrt{\hat{\rho}}\hat{\tau}\sqrt{\hat{\rho}}} \right], \end{equation}
is the square root of quantum fidelity between $\hat{\rho}$, and $\hat{\tau}$. The square root of the fidelity $ \sqrt{F(\hat{\rho},\hat{\tau})}$ can be expressed as (\cite{Bengtsson_Zyczkowski_geometry}, Eq.~$9.30$) 
\begin{equation}
   \label{eq:deforiginal}
    \sqrt{F(\hat{\rho},\hat{\tau})} = \max_{\substack{\hat{\rho} =  AA^\dagger \\
      \hat{\tau} = BB^\dagger}
    }\mathrm{Tr}[AB^\dagger + BA^\dagger]/2,
\end{equation}
where the maximization is over the \textit{complex amplitudes} $\{A,B\}$ which constitute a polar decomposition of $\{\hat{\rho},\hat{\tau} \}$ respectively. Eq.~\eqref{eq:deforiginal} is actually the original definition of quantum fidelity motivated by the concept of transition probability. In fact, if both states are pure $\{\hat{\rho}=\ketbra{\psi}, \hat{\tau}= \ketbra{\phi}\} $, then $\{A= \ket{\psi}, B = \ket{\phi}\}$ and the RHS of Eq.~\eqref{eq:deforiginal} amounts to overlap $|\braket{\psi}{\phi}|^2$. Note that the decomposition admits a gauge degree of freedom $A\mapsto AU$ for $U$ unitary (and similarly for $B$). Our work resolves redundancy using the \textit{Cholesky decomposition} defined in the main text.   

From Eq.~\eqref{eq:deforiginal} we see that for any polar decomposition (and in particular the Cholesky as canonical one, $\{\hat{\rho} = C_\rho C_\rho^\dagger, \hat{\tau} = C_\tau C_\tau^\dagger\}$ ), the following inequality always holds:
\begin{equation}
   \sqrt{ F(\hat{\rho},\hat{\tau})} \geq  \mathrm{Tr}[C_\rho C_\tau^\dagger +C_\tau C_\rho^\dagger]/2.
\end{equation}
Finally, rewriting $2$ as $1+1 = \mathrm{Tr}(\hat{\rho}+\hat{\tau}) = \mathrm{Tr}(C_\rho C_\rho^\dagger+C_\tau C_\tau^\dagger)$ we arrive at
\begin{equation}
\begin{split}
    D_{\mathrm{B}}(\hat{\rho},\hat{\tau})\leq & \mathrm{Tr}[C_\rho C_\rho^\dagger+C_\tau C_\tau^\dagger-C_\rho C_\tau^\dagger-C_\tau C_\rho^\dagger] \\
    =& \mathrm{Tr}[(C_\rho -C_\tau)(C_\rho-C_\tau)^\dagger] \\ =&  D_{\mathrm{HS}}^2(C_\rho,C_\tau),
\end{split}
\end{equation}
where the HS distance defined in the main text is extended to complex matrices (not necessarily Hermitian) as: 
\begin{equation}
\label{eq:HSdistcomplex}
    D_{\mathrm{HS}}^2(C_\rho, C_\tau) = \mathrm{Tr}[(C_\rho- C_\tau)(C_\rho-C_\tau)^\dagger].
\end{equation}\\ 

\noindent\textbf{Hilbert-Schmidt distance as mean squared error (MSE)}.-- In the following we connect the Hilbert-Schmidt distance [Eq.~\eqref{eq:HSdistcomplex}] between two Cholesky matrices $\{C_\rho, C_\tau\}$ associated to quantum states $\{\hat{\rho},\hat{\tau} \}$, with the mean-squared error of the matrix elements.

First, consider a $d\times d$ complex matrix $K$, $\{ K_{\alpha\beta}\}_{\alpha,\beta\in [d]\times [d]}$.
Next, let us introduce the vectorization $\vec{K}$ of its matrix elements as
\begin{equation}
    \vec{K} = \Re{\mathbb{K}}\oplus\Im{\mathbb{K}}\ ,
\end{equation}
where $\mathbb{K}$ is the flattening of the matrix, i.e., $\mathbb{K} = (K_{11},K_{12},..,K_{dd})$, and $\oplus$ the direct sum of vectors, $\vec{v}\oplus\vec{u} = (\vec{v},\vec{u})$. Since $\mathrm{diag}(K)\in \mathbb{R}$ is sufficient for neural network output, the length of the real vector $\vec{K}$ is $d^2$ [$d$ for the diagonal and $d(d-1)/2\times 2$ for the lower triangle ($\times 2$ for the real and imaginary part); the remaining elements are zero].

Let $K = C_\rho - C_\tau$, then the square HS distance, Eq.~\eqref{eq:HSdistcomplex}, reads
\begin{align}
D^2_{\rm HS}(C_\rho,C_\tau) &= \mathrm{Tr}(KK^\dagger) =\sum_{\alpha,\beta}|K_{\alpha\beta}|^2 \\
&= \sum_{\alpha,\beta}\Re{K_{\alpha\beta}}^2+\Im{K_{\alpha\beta}}^2 \\
&= (\Re{\mathbb{K}}\oplus\Im{\mathbb{K}})\cdot (\Re{\mathbb{K}}\oplus\Im{\mathbb{K}}) \\
&= ||\vec{C}_\rho - \vec{C}_\tau||^2  = \mathrm{MSE}(\vec{C}_\rho,\vec{C}_\tau),
\end{align}
Finally, we observe that the MSE is the natural cost function of a standard feed-forward neural network. \\

\section{Architecture details}\label{app: architecture details}

The first layer $h_\mathrm{cnn}$ applies a set of $K$
fixed-size trainable one-dimensional convolutional kernels
to $\vec{C}_\rho$ followed by a non-linear activation function, i.e. $\gamma(h_\mathrm{cnn}(\vec{C}_\rho))\to \{  \boldsymbol{F}^1_\mathrm{cnn}, \dots, \boldsymbol{F}^{K}_\mathrm{cnn}\}$. During the training process, the convolutional kernels learn different features of the dataset, which are then fed to the transformer block $h_{\rm tr}$.
The transformer block $h_{\rm tr}$ distills the correlations between the features extracted from the kernels through the self-attention mechanism, providing a new set of vectors, that is, $h_{\rm tr}(\boldsymbol{F}^1_\mathrm{cnn}, \dots, \boldsymbol{F}^{K}_\mathrm{cnn}) \to \{  \boldsymbol{F}^1_\mathrm{tr}, \dots, \boldsymbol{F}^{K}_\mathrm{tr}  \}$.
In the last step, the outputs of the convolutional kernel of the layer $h_\mathrm{cnn}$ are added and form an output $\vec{C}_{\boldsymbol{\theta}}$, $\mathrm{tanh}(h_\mathrm{cnn}(\boldsymbol{F}^1_{\rm tr}, \dots,\boldsymbol{F}^{K}_{\rm tr})) \to \vec{C}_{\boldsymbol{\theta}}$.
The output is combined in ${C}_{\boldsymbol{\theta}}$ as in Eq.~\eqref{eq:inference}, and a custom cost function $\eta$ is used to apply the RElU activation on the diagonal elements of the reconstructed matrix and the $\tanh$ on the off-diagonal elements. The role of the attention mechanism is explored in Appendix~\ref{app:benchmarking}, where we benchmark the transformer block against CNNs.

The training data and the considered architecture allow interpreting the trained neural network as a conditional debiaser (for details, see the Appendix~\ref{appendix:debiaser}). Although the proposed protocol cannot improve the predictions of unbiased estimators, any estimator that outputs valid quantum states (e.g., LI, MLE) must be biased due to boundary effects. In the given framework, the task of the neural network is to learn such skewness and drift the distribution towards the true mean.  
\vspace{5pt}
\\

\noindent\textbf{Computational details.--} The architecture is trained on 10000 data in training and 1500 in validation, with a batch size of 1500. The code is run on a Nvidia A100 GPU card with 80GB of memory (Cuda version 12.1). The total training time amounts to $\sim25$ minutes. A similar time can be obtained on a commercially available GPU virtual machine.

\section{Sampling random states}
\label{app:random}

We want the algorithm to learn how noise affects generic states that cover the whole space $\mathcal{S}$ of proper quantum states. In doing so, we will have a flexible solution that can potentially improve any state. To this end, we will train our neural network with random states sampled from different measures that uniformly cover the volume of interest. In particular, for mixed states, we will use the Hilbert-Schmidt ensemble, while for pure states, the Haar measure; in the following, we review how to generate such random states:      \\

\noindent\textbf{Hilbert-Schmidt states}.-- The HS measure is defined by the infinitesimal line element: 
\begin{equation}
   \label{eq:HS_line}
    ds_{\mathrm{HS}}^2 = \mathrm{Tr}[(d\hat{\tau}_{\mathrm{HS}})^2],
\end{equation}
which indeed induces the HS-distance $D_{\mathrm{HS}}$, as defined in the main text, once integrated. 

States uniformly distributed in the space endowed with the corresponding metric, Eq.~\eqref{eq:HS_line}, may be sampled as:  
\begin{equation}
\hat{\tau}_{\mathrm{HS}} = \frac{AA^\dagger}{\mathrm{Tr}[AA^\dagger]},
\end{equation}
where $A$ is a generic complex square matrix whose real and imaginary matrix elements are i.i.d. sampled normal random variables $\mathcal{N}(0,1)$. \\

\noindent\textbf{Haar vectors}.-- Such ensemble is defined on the pure state space $\hat{\tau} = \ketbra{\Psi}$ and may be used to infer states with high purity, as the metrologically useful states explored in the main text. Formally, it takes advantage of the link between states $\ket{\Psi}\in \mathcal{H}$, and the unitary group. The fundamental property of such measure is that it is invariant under unitary transformations, thus making any vector in $\mathcal{H}$ equiprobable.

Interestingly, Haar vectors (or equivalently unitaries), can be drawn again from i.i.d. random normal variables $\mathcal{N}(0,1)$ \cite{Mezzadri2006}. Let $A$ be a complex square matrix sampled as before. Then, we apply the so-called QR decomposition: 
\begin{equation}
    A = QR,
\end{equation}
where $Q$ is unitary and $R$ is upper-triangular. Then $Q' = Q\Lambda $, where $\Lambda = \mathrm{diag}(\{R_{\alpha\alpha}/|R_{\alpha\alpha}| \}_\alpha)$, is Haar random. \\

In practice, we use the python package \textit{qutip} to compute those. For further information on random states, we refer to the book Ref.~\cite{Bengtsson_Zyczkowski_geometry}.

\section{Capturing the metrological usefulness of OAT evolved states}
\label{sec: app_QFI}

Given a quantum state $\hat{\rho}$ with spectral decomposition $\hat{\rho}=\sum_k p_k\ketbra{k}$, we can evaluate its sensitivity upon rotations generated by $\hat{G}$, $\hat{\rho}(\theta) =e^{-i\theta \hat{G}}\hat{\rho}e^{+i\theta \hat{G}} $ as quantified by the QFI: 
\begin{equation}
 F_{Q}[ \hat{\rho}, \hat{G}] = 2 \sum_{\substack{k,l \\ p_k +p_l >0}}\frac{(p_k -p_l )^2 }{p_k + p_l } | \bra{k}\hat{G}\ket{l}|^2 .
 \label{eq:def_QFI}
\end{equation}
In the main text, we consider an ensemble of $L$ spins-$1/2$, or two-level atoms. In such system and in the context of magnetometry (e.g., via Ramsey interferometry), the phase is encoded the same way in every constituent via the collective generator $\hat{J}_{\mathbf{v}}=\sum_{i\in [L]}\mathbf{v}\cdot\hat{\boldsymbol{\sigma}}_i$ where $\mathbf{v} = (v_x,v_y,v_z)$ is the encoding orientation. 

For a generic state $\hat{\rho}$, it is not direct to find the optimal spatial direction $\mathbf{v}$ to exploit the maximal sensitivity. However, for pure states $\hat{\rho} = \ketbra{\Psi}$ (like the OAT evolution that we consider), the QFI is (four times) the variance of the generator, $F_{Q}[\ket{\Psi}, \hat{G}] = 4[\bra{\Psi}\hat{G}^2\ket{\Psi} - \bra{\Psi}\hat{G}\ket{\Psi}^2]$ and the best direction is then yielded by the maximal eigenvalue $\mathbf{v}_{\rm max}$ of the covariance matrix $\mathcal{C}$:  
\begin{align}
    \mathcal{C} =& \begin{pmatrix}
\langle \hat{J}_x^2 \rangle & \mathrm{Re}\{\langle \hat{J}_x\hat{J}_y \rangle\} & \mathrm{Re}\{\langle \hat{J}_x\hat{J}_z \rangle\} \\
\mathrm{Re}\{\langle \hat{J}_x\hat{J}_y \rangle\} & \langle \hat{J}_y^2 \rangle & \mathrm{Re}\{\langle \hat{J}_yJ_z \rangle\} \\
\mathrm{Re}\{\langle \hat{J}_x\hat{J}_z \rangle\} &  \mathrm{Re}\{\langle \hat{J}_y\hat{J}_z \rangle\} & \langle \hat{J}_z^2 \rangle
    \end{pmatrix}  \\
    &- 
    \nonumber
    \begin{pmatrix}
        \langle \hat{J}_x \rangle\\
        \langle \hat{J}_y \rangle\\
        \langle \hat{J}_z \rangle
    \end{pmatrix}
    \begin{pmatrix}
         \langle \hat{J}_x \rangle &  \langle \hat{J}_y \rangle &  \langle \hat{J}_z \rangle
    \end{pmatrix},
\end{align}
where the expectation value is taken against pure states, $\langle\cdot \rangle:=\bra{\Psi}\cdot\ket{\Psi}$. 
The maximal value of the QFI achieved is consequently $F_{Q}[\ket{\Psi}, 
J_{\mathbf{v}_{\mathrm{best}} = \mathbf{v}_{\max}}]= 4\mathbf{v}_{\max}^T\mathcal{C}\mathbf{v}_{\max} = 4\lambda_{\max}(\mathcal{C})$, where $\lambda_{\max}$ indicates maximal eigenvalue. Here, we will introduce two basic examples of such results: 
\begin{itemize}
    \item A coherent spin state pointing along $x$ of length $J = L/2$, $\ket{+}^{\otimes L}$. This initial state was chosen to start OAT dynamics, Eq.~\eqref{eq:def_OAT} (main text). In such case, the optimal axis is any orientation orthogonal to $x$, i.e., contained in the $yz$ plane. The QFI achieved is exactly $L$, which is the maximal value that can be reached by separable states (a.k.a. shot noise limit) \cite{pezze2009entanglement}.   
    \item The GHZ or cat state aligned along $x$, $(|+\rangle^{\otimes L}+ e^{i\phi}|-\rangle^{\otimes L})/\sqrt{2}$, which is realized at time $t=\pi/2$ during the OAT dynamics. Now, the optimal generator points in the $x$ direction, and the corresponding QFI (variance times four) is $L^2$. Such value actually is the maximal QFI achievable within the quantum framework and requires genuine $L$-partite entanglement \cite{hyllus2012fisher,toth2012multipartite}.  
\end{itemize}
Since the OAT reconstructed states $\hat{\rho}$ are of high purity, the same procedure can approximate the optimal orientation $\mathbf{v}$. However, the QFI results are evaluated exactly as per Eq.~\eqref{eq:def_QFI}. For further study, we refer the reader to the excellent review of Ref.~\cite{RevModPhys.90.035005}.

\section{OAT state analysis for zero channel and zero measurement noise}\label{OAT ideal}

In this section, we consider an optimal scenario, when only statistical sampling noise is being considered. To prepare our data, we also consider SIC-POVM operators this time, given their higher performance in tomography reconstruction tasks.
The quality of reconstruction is shown in Table~\ref{tab:nn only sampling}.
First, we verify that the neural network can improve substantially for the OAT states, even though no examples of such states were given in the training phase, which relied only on Haar-random states.
Moreover, the OAT-averaged reconstruction values exceed the Haar reconstruction ones. 
We conjecture that this stems from the bosonic symmetry exhibited by the OAT states. 
This symmetry introduces redundancies in the density matrix, which might help the neural network detect errors produced by statistical noise. 
Finally, let us highlight that the network also displays good robustness to noise.
In fact, when we feed the same network with states prepared for $N_{\rm trial} = 10^3$ trials, we increase the fidelity of the reconstruction from $67\%$ to $87\%$. \\

\begin{table}[h!]
\centering 
    \begin{tabular}{|c||p{1.5cm}|p{1.5cm}|p{1.5cm}|p{1.5cm}|}
    \hline
    &$10^6$ trials &$10^5$ trials & $10^4$ trials& $10^3$ trials\\ 
          \hline
          LI-OAT &$98.1 \pm 0.3$ & $94.3\pm 0.4$ & $86.5\pm 0.1$& $67.0\pm 2.0$\\

    \hline 
      NN-OAT & $99.3\pm 0.2$ & $98.6 \pm 0.4$ & $97.8 \pm 0.5$  & $87.6\pm 4.1$\\
      \hline 
   NN-Haar   &$99.0 \pm 0.2$ & $96.9\pm 3.0 $ & $94.2 \pm 3.3$& $81.1 \pm 4.1$\\
     \hline
  
    \end{tabular}
    \caption{Comparison of average fidelity and its standard deviation between the reconstructed and the target states of size $d=16$ for various QST methods (rows), with varying size of measurement trials $N_{\rm trial} = 10^6, 10^5, 10^4, 10^3$, as indicated by the consecutive columns.    
    The first row presents the average fidelity reconstruction for linear inversion QST, averaged over OAT states, evenly sampled from $t = 0$ to $t = \pi$.
    Employing our neural network presents an enhancement over the bare LI, as shown in the second row for the same target set.
    Finally, the third row also shows data for NN-enhanced LI but averaged over general Haar-random states. All initial Born values are calculated by noiseless SIC-POVM.}
    %\label{sic table}
    \label{tab:nn only sampling}
\end{table}

\begin{figure}[t!]
\includegraphics[width=\linewidth, right]{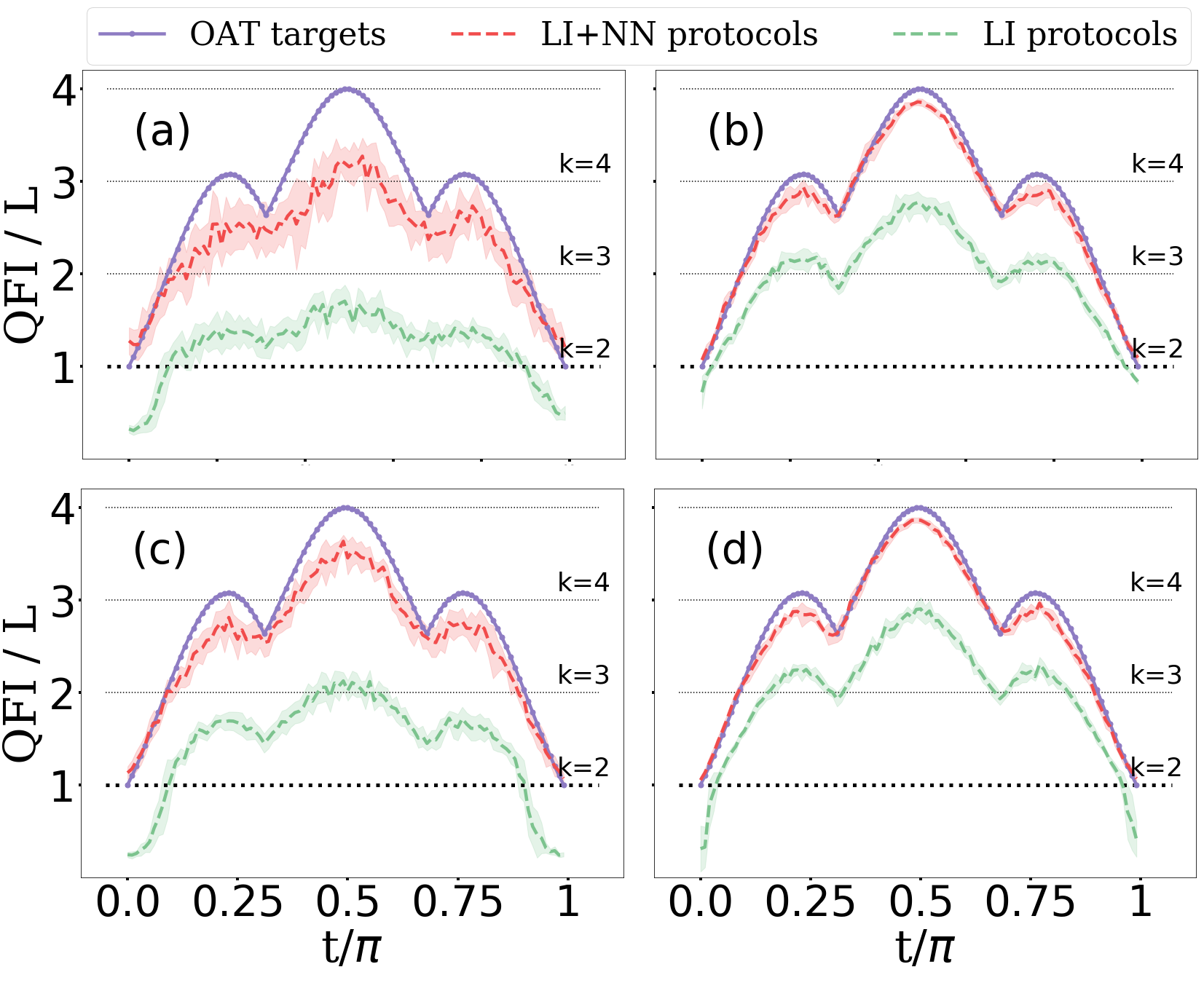}
\caption{Time evolution of the normalized QFI 
during the OAT protocol for $L=4$ qubits system.
Solid blue lines represent QFI calculated for target quantum states. The mean values of QFI calculated from tomographically reconstructed density matrices are denoted by green-dashed (reconstruction via LI), and red-dotted lines (reconstruction via neural network post-processed LI outputs). Shaded areas mark one standard deviation after averaging over $10$ reconstructions. Panels (a) and (b) correspond to LI protocol with SIC-POVM data, whereas (c) and (d) denote LI reconstruction inferred from Pauli measurements. 
In the upper row, the left (right) column corresponds to $N_{\rm trial} = 10^3$ ($10^4$) trials; in the lower row, the left (right) column reproduces the LI initial fidelity reconstruction of $\sim 74\% (\sim 86\%)$. 
The red lines represent the whole setup with neural network post-processing of data from corresponding green lines, indicating improvement over the LI method. 
The neural network advantage over the bare LI method can be characterized by entanglement depth certification, as shown by the horizontal lines denoting the entanglement depth bounds ranging from the separable limit (bottom line, bold) to the genuine $L$-body limit (top line).
In particular, the presence of entanglement, $k\geq 2$, is witnessed by $\mathrm{QFI}>L$, as shown by the violation of the separable bound (bold horizontal line).}
\label{fig:OAT}
\end{figure}

\section{Benchmarking the transformer layer}
\label{app:benchmarking}

The transformer architecture represented a breakthrough in deep learning for sequential learning problems, such as natural language processing~\cite{attention}. 
It has been used for QST problems that involve an unsupervised approach based on the POVM ansatz~\cite{Cha2021}, and more complex architectures to extract diffusion parameters of Brownian motion~\cite{Requena2022}. 
Here, we show the usefulness of the transformer architecture in improving our QST \textit{matrix-to-matrix} protocol. 
This becomes especially pronounced in the regime of highly correlated, that is, entangled quantum states, as shown in the latter part of this appendix.
To test its role, we explore and compare our transformer model with two different convolutional-only architectures (CNN) based on their reconstruction ability. In particular, we benchmark against two setups: (i) a four-layer convolutional neural network, where the transformer layer is replaced with two convolutional ones,
\begin{equation}
    l_{\boldsymbol{\theta}}^{(4)}[\vec{C}_{\rho}] = \Big( \gamma (h_{\mathrm{\mathrm{cnn}}})\Big) ^{\circ 4} [\vec{C}_{\rho}]
\end{equation}
and (ii) the simplest CNN model consisting of two convolutional layers, represented as,
\begin{equation}
    l_{\boldsymbol{\theta}}^{(2)}[\vec{C}_{\rho}] =  \Big(\gamma(h_{\mathrm{\mathrm{cnn}}}) \Big)^{\circ 2} [\vec{C}_{\rho}], \ 
\end{equation}
where $\gamma$ is the GELU activation function in both the cases.

% Formally, the architecture selected for the benchmark are}:
% \begin{align}
%  l_{\boldsymbol{\theta}}^{(4)}(\vec{C}_{\rho}) =& \gamma [h_{\mathrm{\mathrm{cnn}}}\circ\gamma(h_{\mathrm{\mathrm{cnn}}})\circ\gamma(h_{\mathrm{\mathrm{cnn}}})\circ \gamma(h_{\mathrm{\mathrm{cnn}}})  ](\vec{C}_{\rho})\\
%   l_{\boldsymbol{\theta}}^{(2)}(\vec{C}_{\rho}) =& \gamma[h_{\mathrm{\mathrm{cnn}}}\circ \gamma(h_{\mathrm{\mathrm{cnn}}})  ](\vec{C}_{\rho}) \ ,
% \end{align}
We set an equivalent number of trainable parameters for all three architectures. We tested the two CNNs for the same datasets as previously used for mixed- and pure-state reconstruction. In the following, we analyze the performance of the models for mixed- and pure-state reconstruction. \\

\begin{figure}[t!]
    \centering
    \includegraphics[width = .48\textwidth]{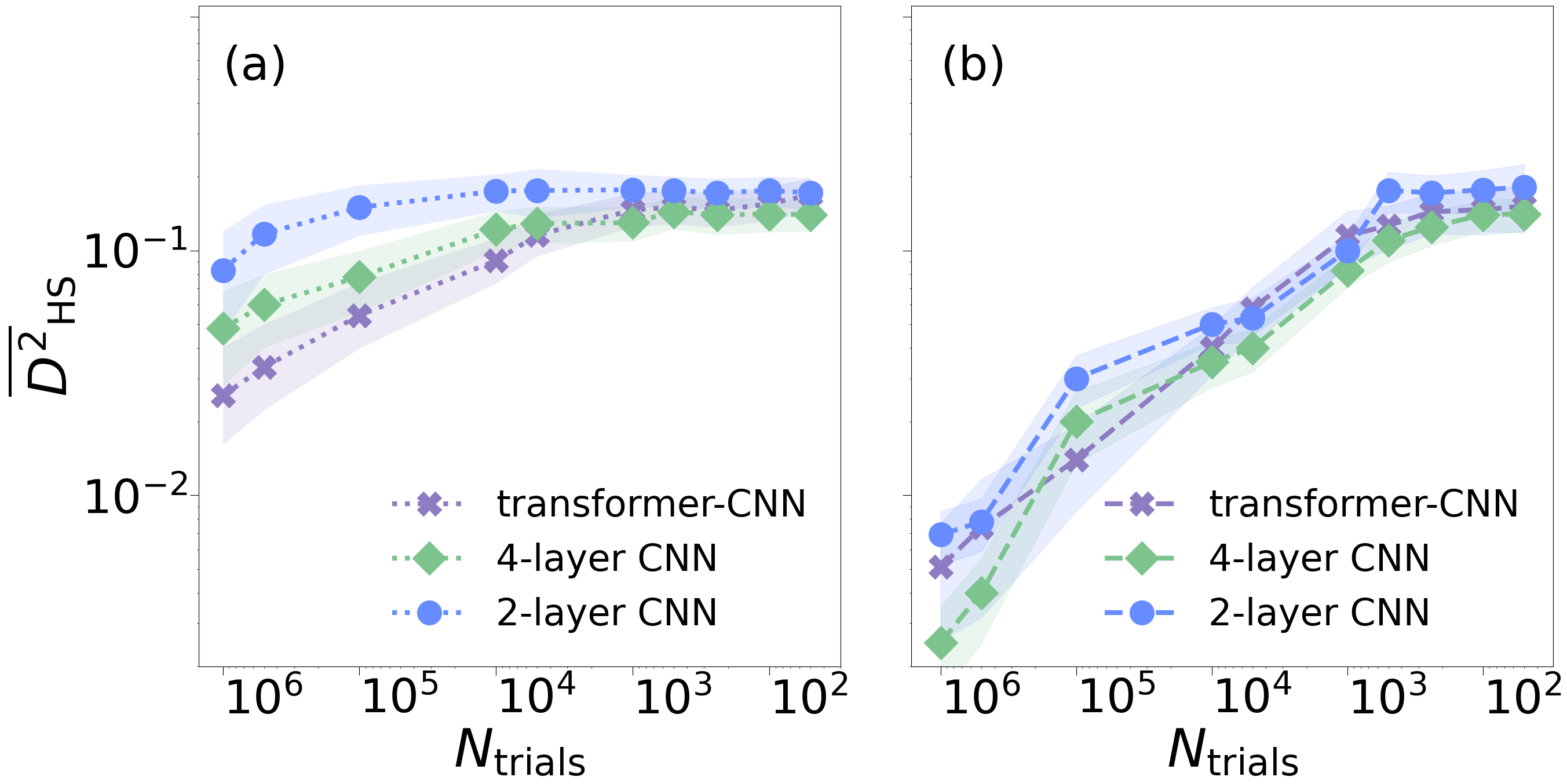}
    \caption{Comparison of the efficiency of QST reconstruction schemes evaluated using Hilbert-Schmidt distance square $\overline{D^2}_{\rm HS}$ for transformer-based, 2-layer, and 4-layer attention-free CNN models, averaged in 1000 mixed states. All models share an equivalent number of training parameters. (a) Average reconstruction values for the 10 different LI pre-processed test datasets. Similarly to Fig.~\ref{fig: benchmark}, we vary the number of trials $N_\text{trials}$ to analyze the reconstruction efficiency and also use states of dimension 9 for a direct comparison.
    (b) The same analysis applied to the models trained on the MLE pre-processed data. To summarize, only for the MLE pre-processed data, the 4-layer CNN model can outperform the transformer-based for $N_{\rm trials} = 10^6,10^5$, while for the LI pre-processed our network shows better outcomes. }    
    \label{fig: cnn-benchmark}
\end{figure}

\noindent{\bf Mixed state reconstruction.--}
In Fig.~\ref{fig: cnn-benchmark}, we show the reconstruction of mixed states using the three setups, namely our transformer and the two CNN architectures. Firstly, as shown in panel (a),  for the LI pre-processed data, our transformer-based model outperforms CNNs by showing a higher expressivity in terms of better reconstruction ability for a large number of trials $N_{\text{trials}}>10^4$.
However, the three models are almost equivalent in the undersampled regime. Next, in panel (b), we show our numerical experiments conducted on the MLE pre-processed dataset, which has a comparatively lesser amount of noise. We can see that the three models are almost equivalent.\\

\noindent{\bf Pure state reconstruction.--} 
For Haar-random pure states, using any of the two CNN models for the QST protocol causes an evident drop of $\sim 10\%$ in the fidelity of reconstruction compared to the our attention-based model for the undersampled regime ($N_{\rm trial} = 10^3$), as exemplified in Table~\ref{table two}.

\begin{table}[h!] 
\centering 
    \begin{tabular}{|c||p{2cm}|p{2cm}|p{2cm}|}
    \hline
    &$10^5$ trials & $10^4$ trials& $10^3$ trials\\ 
          \hline
          CNNs  & $94.2\pm 3.0\%$ & $87\pm 2.2\%$ & $68.0\pm 4.0\%$\\
          \hline 
   Transformer    & $96.9\pm 3.0 \%$ & $94.2 \pm 3.3\%$& $81.1 \pm 4.1\%$\\
  \hline
    \end{tabular}
    \caption{Values of averaged fidelity and its standard deviation between the reconstructed and the target Haar pure states of size $d=16$ obtained by the two CNN architectures $l_{\boldsymbol{\theta}}^{(2)},l_{\boldsymbol{\theta}}^{(4)}$ used inside our matrix-to-matrix protocol. In the second row, the outcomes we obtained by applying the transformed-based model. Similar to Fig.~\ref{fig:OAT}, we check for a different number of trials $N_{\rm trial} = 10^5, 10^4, 10^3$ to analyse the performance of the different architectures. All the initial Born values are calculated via noiseless SIC-POVM.}
    \label{table two}
\end{table}

\begin{figure}
    \centering
    \includegraphics[scale = 0.14]{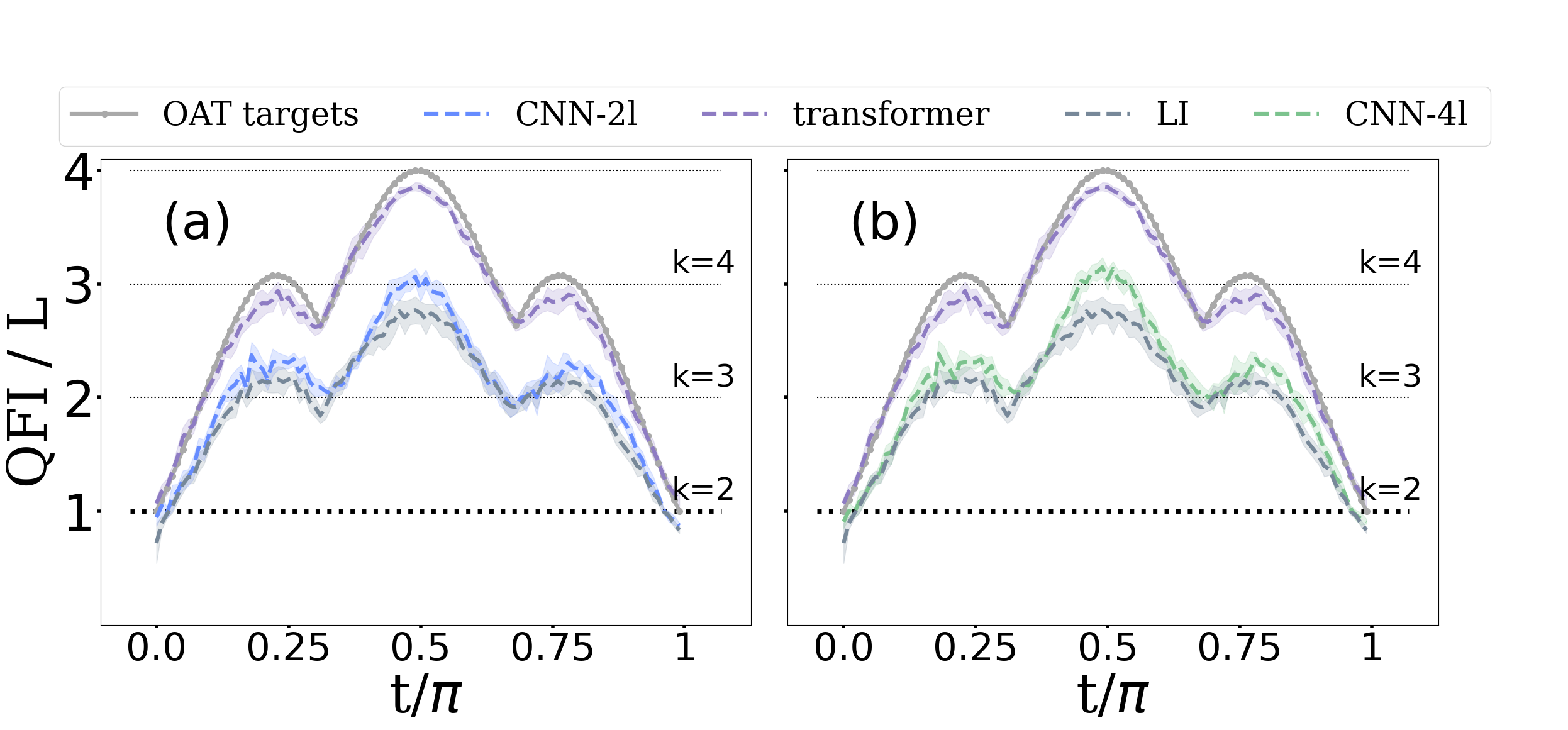}
    \caption{Time evolution of the normalized QFI during the OAT protocol for 4 qubits. The dotted dark grey line represents the QFI calculated for the target quantum state, and the light grey dashed line is the QFI upon LI reconstruction (our minimal threshold). Panels (a) and (b) correspond to the QFI obtained for the states reconstructed by the $2-$layers and $4$-layers CNN respectively. We observe that, firstly, the transformer-based model outperforms the CNN models at all times with reconstruction ability very close to the OAT target states. Secondly, the CNN models perform equivalently irrespective of the number of layers in the architecture as shown in panels (a) and (b) for $2-$layers and $4-$layers respectively when considering QFI as our reconstruction metric.
    }
    \label{fig: pure states benchmark}
\end{figure}

Finally, we apply our protocol to the pure state reconstruction task generated from one-axis twisting dynamics. In Fig.~\ref{fig: pure states benchmark} we present the quantum Fisher information extracted from the reconstructed states. A significant drop in quantum Fisher information reconstruction is observed in using CNN-based architectures compared to the transformer architecture.

We conclude that our transformer-based architecture works significantly well compared to the only CNN-based architectures in the case of reconstructing pure OAT states, when the QFI is considered a figure of merit. In the other case of mixed states reconstruction, the three architectures are comparable, and the transformer-based approach works better in the LI-preprocessed states in the highly sampled regime.

\section{Improving the averaged MLE in HS metric}
\label{app:MLE_and_HS_metric}

The MLE is efficient asymptotically, with a number of measurements $N\to \infty$. However, for a few measurements, there is no guarantee that MLE performs best. Such a situation takes place in the undersampled regime. In terms of the HS metric, it is usually more convenient to ignore any knowledge and just take the maximally mixed state $\mathbb{I}/d$ as an estimation. Here, we propose a simple method to interpolate between the two extreme results by depolarizing the MLE state $\hat{\rho}$,              
\begin{equation}
\label{eq:mix_MLE}
    \hat{\rho}_p =  p\hat{\rho} + (1-p)\mathbb{I}/d \ ,
\end{equation}
with $0\leq p\leq 1$. The parameter $p$ would then incorporate the statistical noise inherent to have a finite number of samples.

\begin{figure}[t!]
    \centering
    \includegraphics[width = .5\textwidth]{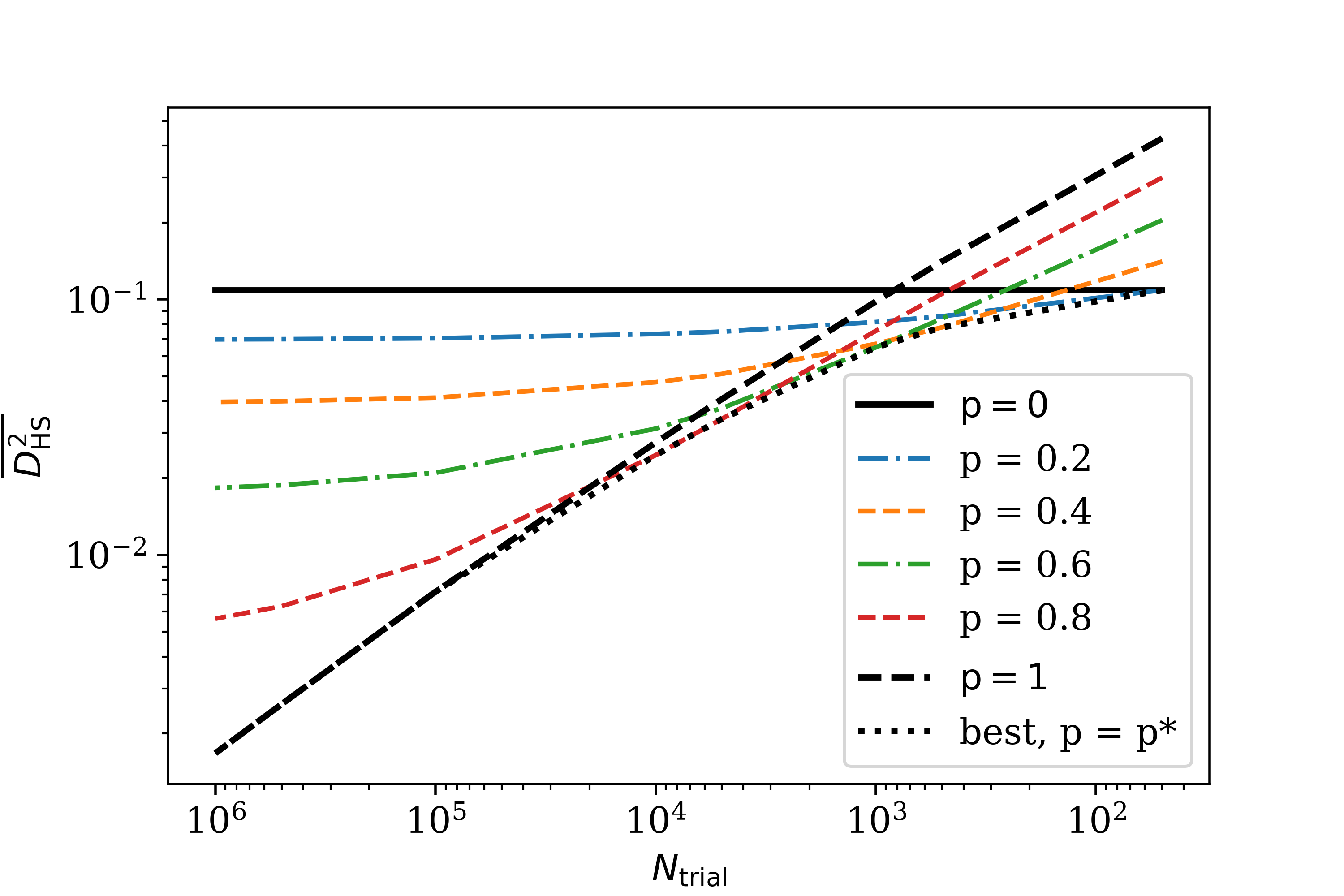}
    \caption{Averaged HS distance of reconstructed MLE from the HS ensemble with $d=9$ as mixed according to Eq.~ \eqref{eq:mix_MLE} for different values of $p$ (coloured lines). We highlight the limiting cases, namely $p=0$ (solid line); the average with respect $\mathbb{I}/d$, and $p=1$ (dashed): the MLE result. The envelope of such a family of lines is marked with a dotted line. Such bound can be realized with an optimal $p_*$, which depends on the number of trials via the reconstructed $\{\hat{\rho}_{\mathrm{MLE}}\}$.}
    \label{fig:interpol}
\end{figure}

In Fig.~\ref{fig:interpol} we show the average HS distance as a function of a number of trials $N_{\rm trial}$, for different values of the parameter $p$. The first extreme case ($p=0$) i.e., maximally mixed state is presented as a horizontal solid line, while the second extreme case ($p=1$), i.e., the MLE reconstructed state, is depicted as a dashed line. All intermediate values of $p$ form an envelope shape, corresponding to a critical value $p^*$, outperforming all other values of $p$.

To calculate the critical $p^*$ let us notice that the average of the squared HS distance with respect to the state Eq.~\eqref{eq:mix_MLE}, $D_p^2 = D_{\mathrm{HS}}^2(\hat{\tau},\hat{\rho}_p)$, can be expressed as,
\begin{equation}
    \label{eq:D_HS_p}
    D_p^2 = D_0^2 + (D_1^2 - D_0^2 - D_{01}^2)p + D_{01}^2p^2 \ ,
\end{equation}
where
\begin{align}
    D_0^2 &=  D^2_{\mathrm{HS}}(\hat{\tau},\mathbb{I}/d) \\
    D_1^2 &=  D^2_{\mathrm{HS}}(\hat{\tau},\hat{\rho}_{\mathrm{MLE}}) \\
    D_{01}^2 &= D^2_{\mathrm{HS}}(\mathbb{I}/d,\hat{\rho}_{\mathrm{MLE}})
\end{align}

\begin{figure}[ht]
    \centering
    \includegraphics[width = .24\textwidth]{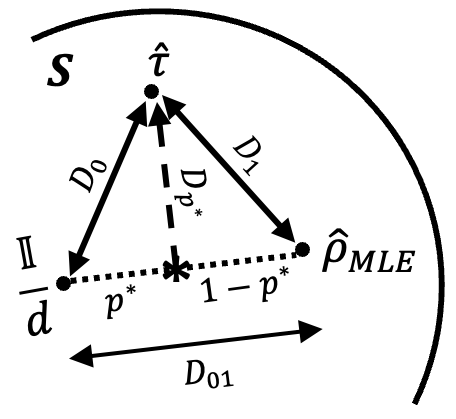}
    \caption{Geometric interpretation of the optimal depolarization of the MLE state, such as incorporating the statistical noise stemming from a finite number of experimental runs.}
    \label{fig:depol}
\end{figure}

The Eq.~\eqref{eq:D_HS_p} is equally valid in average as it is linear in the $D^2$'s, and we consider $p$ a global parameter. Then, $\overline{D_0^2}$ can be related to the average purity $\overline{\mathrm{Tr}(\hat{\tau}^2)}$ as $D_0^2 = \mathrm{Tr}(\hat{\tau}^2)-d^{-1}$. Also, as depicted in Fig.~\ref{fig:depol}, it can be interpreted geometrically as a triangle of sides' length $\{D_0, D_1, D_{01}\}$ in which we want to find the point $p$ in the segment $01$ which is closer to the opposed vertex. Minimization of the same equation leads to the optimal $p^*$:
\begin{equation}
    p_* = \begin{cases}
    0 & \overline{D_{01}^2}< \overline{D_1^2-D_0^2}  \\
    \frac{\overline{D_{01}^2+D_0^2 -D_1^2}}{2\overline{D_{01}^2}}  & \mbox{otherwise}\\ 
    1 &  \overline{D_{01}^2} < \overline{D_0^2-D_1^2} \\
    \end{cases} \ ,
\end{equation}
yielding an optimal distance: 
\begin{equation}
    \overline{D_*^2} = \begin{cases}
    \overline{D_0^2} & \overline{D_{01}^2}< \overline{D_1^2-D_0^2}  \\
    \overline{D_0^2}-\frac{\overline{D_0^2-D_1^2-D_{01}^2}^2}{4\overline{D_{01}^2}}   & \mbox{otherwise}\\ 
    \overline{D_1^2} &  \overline{D_{01}^2} < \overline{D_0^2-D_1^2} \\
    \end{cases} \ ,
\end{equation}

\begin{figure}[ht]
    \centering    
    \includegraphics[width = .5\textwidth]{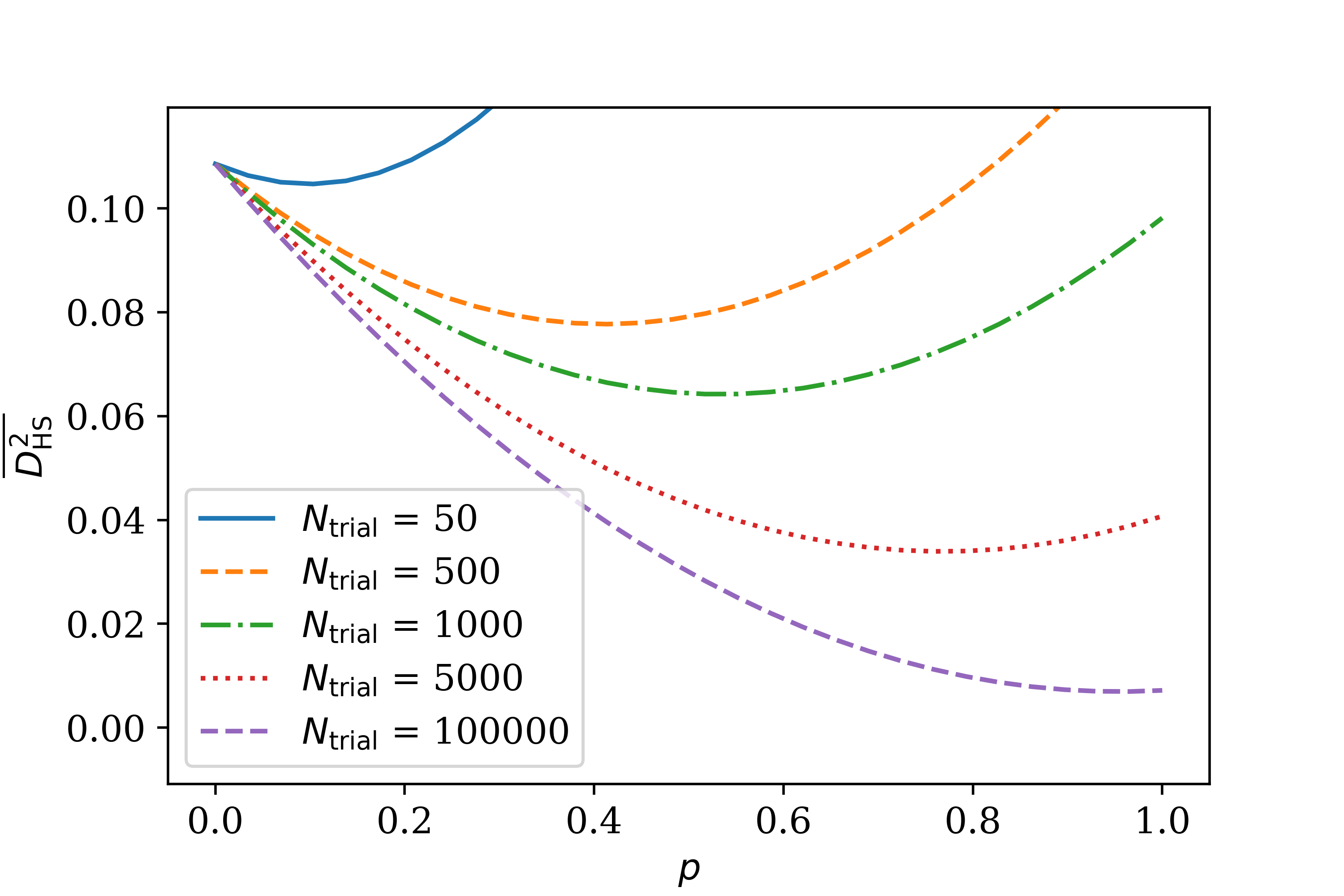}
    \caption{Average reconstruction distance as a function of the mixing parameter $p$ for a given set of number of trials. We verify the parabola curves of Eq.~\eqref{eq:D_HS_p} and the nontrivial minima. }
    \label{fig:parabs}
\end{figure}

In Fig.~\ref{fig:parabs}, we observe a realization of the non-trivial solution (otherwise case), which is better than both $\overline{D_0^2}$ and $\overline{D_1^2}$. \\

Finally, one should check how this method relates to Bayesian approaches \cite{Granade2016}. In particular, how to incorporate partial information about the ensemble, e.g., by only assuming as prior knowledge an average purity.

\section{Interpretation of the neural network as a conditional ``debiaser"}\label{appendix:debiaser}

Our neural network takes as input a reconstructed state $\hat{\rho}$ from the experimental results $\mathbf{f}$ through some estimator (e.g. LI, MLE), $Q[\mathbf{f}]=\hat{\rho}$. The neural network returns a state $\hat{\bar{\rho}}$ that, on average over the realizations (in $\mathbf{f}$)
better approximates the target $\hat{\tau}$ than $\hat{\rho}$.
In the following, we outline the situations in which we expect a poor performance of our algorithm. From the above setting, we immediately see that it is useless for the unbiased estimators. \\

\noindent\textbf{Observation 1.} If the estimator $Q$ is unbiased, i.e., $\hat{\rho} = \hat{\tau}$, no further improvement can be achieved with our approach. In fact, the mean already provides the best estimation. Consequently, if some enhancement is observed, the inference of the input state must be biased. The bias here comes from the requirement that $\hat{\rho}$ must be a proper quantum state. 

To observe why it is so, let us focus on the simplest case of two projective quantum measurements, i.e., spin measurements of an electron in the $x$ and $y$ directions. 
Although both measurements belong to the set $[-1/2,1/2]$, not all pairs of measurement results are admissible. 
For example, there are no valid quantum states for which both measurements yield $1/2$ since then the total spin would be larger than $1/2$.

Therefore, if we account for the inevitable noise present for finite statistics, the unphysicality of certain real-world measurements is unavoidable.
There are two general QST strategies to overcome this obstacle -- either discard the unphysical outcomes (e.g., MLE) or keep them all (e.g., LI). 
Any of these methods has its drawbacks, and, for finite statistics, one cannot have a strategy that satisfies both the linearity and the physicality of the predicted states~\cite{Silva_2017}. \\

\noindent\textbf{Observation 2.} Any estimator $Q$ which, as required by our neural network architecture, outputs a valid state $\hat{\rho}_{\mathbf{f}}$ for any physical $\mathbf{f}\in \mathcal{F}$, is biased.  \\

\begin{figure}[h!]
    \centering
    \includegraphics[width = .4\textwidth]{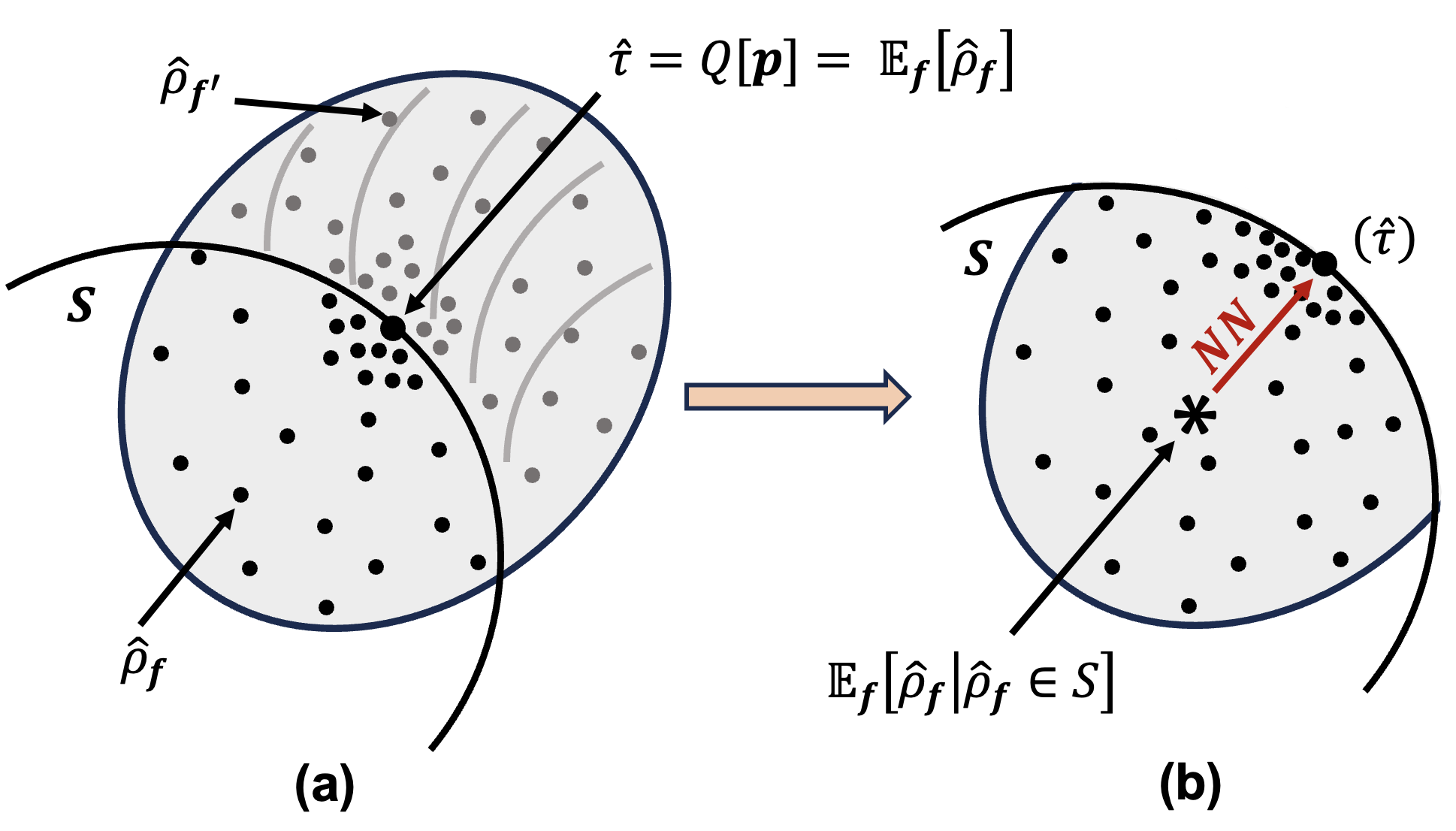}
    \caption{Action of the neural network as a conditional debiaser. (a) Inference of the state $\hat{\tau}$ by many finite-size realization $\{\hat{\rho}_{\mathbf{f}} \}_{\mathbf{f}}$ not necessarily a proper state (i.e., it might be outside $\mathcal{S}$). (b) Disregarding the non-physical realizations results in a skewed conditional distribution whose mean is displaced from the true state. The action of the neural network is then to shift back the mean to the target state by drifting the distribution. }
    \label{fig:debias}
\end{figure}

This phenomenon becomes more prominent for $\hat{\tau}$ close to the boundary of the quantum set $\mathcal{S}$, i.e., with high purity \cite{Ferrie2018,Scholten2018}. By convexity, the boundary of $\mathcal{F}_\mathcal{S}$ is attained by extremal elements of $\mathcal{S}$.
From which we notice that the chance of non-quantum outcomes is higher and $\mathbb{E}_{\mathbf{f}}[\hat{\rho}_{\mathbf{f}}|\hat{\rho}_{\mathbf{f}}\in\mathcal{S}]$ becomes significantly displaced from $\hat{\tau}$ (see sketch Fig.~\ref{fig:debias}). In such terms, the task of the neural network can be interpreted as drifting the skewed probability distribution towards the true mean, by means however of only quantum outcomes. In other words: \\

From $\{\mathbf{f} \in \mathcal{F}_{\mathcal{S}}\}$, find ensemble $\{\mathbf{f}_{\mathrm{NN}} \in \mathcal{F}_{\mathcal{S}} \}$ such that the output bias $|\mathbb{E}_{\mathbf{f}_{\mathrm{NN}}}(\hat{\rho}_{\mathbf{f}_{\mathrm{NN}}})-\hat{\tau}| $ is minimal.  \\

As a result, generic mixed states are harder to improve than pure states. More studies are needed to confirm the said behavior. In particular, one has to verify that no improvement is possible if the reconstruction method to obtain $\hat{\rho}_\mathbf{f}$ already incorporates the skewness and how performance depends on the bias of the estimator $Q$.

\end{document}